\documentclass{aa}  

\usepackage{graphicx}
\usepackage{txfonts}
\usepackage{color}
\usepackage[normalem]{ulem}

\begin{document}

   \title{Stellar expansion or inflation?}


   \author{Gautham N. Sabhahit
          \inst{1}
          \and
          Jorick S. Vink\inst{1}
          }

   \institute{Armagh Observatory and Planetarium, 
              College Hill, Armagh BT61 9DG, N. Ireland\\
              \email{gauthamns96@gmail.com}
             }


 
  \abstract
   {While stellar expansion after core-hydrogen exhaustion related to thermal imbalance has been documented for decades, the physical phenomenon of stellar inflation that occurs close to the Eddington limit has only come to the fore in recent years. We aim to elucidate the differences between these physical mechanisms for stellar radius enlargement, especially given that additional terms such as `bloated' and `puffed-up' stars have been introduced in the recent massive star literature. We employ single and binary star MESA structure and evolution models for constant mass, as well as models allowing the mass to change due to winds or binary interaction. We find cases that were previously attributed to stellar inflation in fact to be due to stellar expansion. We also highlight that while the opposite effect of expansion is contraction, the removal of an inflated zone should not be referred to as contraction but rather \textit{deflation,} as the star is still in thermal balance.}


   \maketitle
%

\section{Introduction}

After stars exhaust their hydrogen (H) fuel in the core, they expand -- a phenomenon commonly observed in stellar models with a typical core-envelope structure. The total radius dramatically increases by nearly two or three orders of magnitude beyond the main sequence (MS). While the exact physical origins of this redwards expansion are still discussed \citep{Sugimoto2000, Stancliffe2009, Hekker2020, Miller2022, Renzini2023}, it is accepted that it is related to thermal imbalance.  


By contrast, more massive stars approach the Eddington limit for radiation pressure, which, ever since the introduction of OPAL opacities, has been shown to lead to a peculiar envelope density profile involving inflated tenuous radiation-dominated layers \citep{Ishi1999, Petrovic2005} on top of a dense convective base \citep[see $\mathrm{Fig.}\,1$ in][]{Graf2012}. Envelope layers near their local Eddington limit inflate to lower densities to prevent a potential super-Eddington scenario within the star. Such inflated layers can arise already during the MS, when the models are in hydrostatic and thermal equilibrium.

While this second physical phenomenon is generally less relevant for low-mass stars, it could be the more relevant physical radius enlarger for high-mass stars. When stellar models show larger radii and lower temperatures, it is not immediately clear whether this supergiant configuration resulted from stellar expansion or inflation (or both). For instance, while \citet{Graf2012} discussed the issue of the variability of luminous blue variables in the context of inflation from the iron (Fe) opacity bump, \citet{Sanyal2015} extended this phenomenology also to the hydrogen and helium (He) bumps, which are located at lower temperatures, i.e., larger radii. However, {it} is not immediately obvious if those envelopes are inflated in the same way as those from the Fe-bump discussed by \citet{Graf2012}, or involve a more classical stellar expansion instead. In fact, even the question of whether massive red supergiants (RSGs) have always been subjected to expansion or if the higher-mass RSGs could be RSGs due to inflation only has not yet been addressed.

{Whether models undergo expansion or inflation (or both) has significant implications for a broad range of objects, including implications related to the evolution and final fates of the most massive stars, the luminosity distribution of cool supergiants, and the formation of heavy black holes (BHs) below the pair-instability boundary, among others.}  
This is especially relevant when considering the temperatures and radii of supergiants in terms of blue or red supergiant configurations, as these are important for understanding the maximal BH masses from stars over a range of metallicities \citep{Vink2021}. Internal mixing processes, such as convective overshoot, could either help push the star over to the red, or, with lower amounts of core overshoot, keep the star blue and compact instead. Again it is not immediately clear from the outset whether the most relevant physics is that of envelope expansion, inflation, or both. 

Finally, there has been a flurry of interesting developments in the binary interaction community, particularly concerning stripped and partially stripped stars that appear cooler in the HR diagram than the helium zero-age main sequence (He-ZAMS) and, at times, even cooler than the hydrogen zero-age main sequence (H-ZAMS) \citep{Gotberg2017, Gilkis2019, Laplace2020, Klencki2020, Klencki2022, Ramachandran2023, Ramachandran2024}. This has led to the introduction of terms in the literature such as `bloated' or `puffed-up' (stars), though the exact physical mechanisms these binary models refer to are not always clearly defined.

The present paper is organised as follows. In $\mathrm{Sect.\,}\ref{sec: MESA_methods},$ we provide an overview of the stellar evolution models used in this study, detailing the input parameters and certain definitions that are crucial in subsequent sections. In $\mathrm{Sect.\,}\ref{sec: exp_vs_infl},$ we distinguish between the phenomena of expansion and inflation. The general properties of expanded and inflated envelopes are outlined, along with the conditions under which a blend of these two morphologies might occur. We then discuss important examples of inflation in the context of specific stages of evolution of massive stars in $\mathrm{Sect.\,}\ref{sec: discussion}$. The role of expansion or inflation processes in stabilising a model as a blue or red supergiant, which could potentially influence the resulting BH mass, is explored in  $\mathrm{Sect.\,}\ref{sec: Heavy_BH}$. We analyse single-star configurations in the post-binary interaction context in $\mathrm{Sect.\,}\ref{sec: binary}$.

\section{MESA modelling}
\label{sec: MESA_methods}

In this section, we provide an overview of the stellar evolution inputs used to generate our models. The models presented in this study are produced using the 1D stellar evolution code MESA (version r12115) \citep{MESA11, MESA13, MESA15, MESA17, MESA19}. 

\subsection{Input parameters}

For single-star models, the initial mass ranges from  $10-200\,M_\odot$, and the initial metal mass fractions are $Z = 0.02$ and $Z = 0.0002$. We also run binary-star models with the initial mass and initial period  fixed at $20\,M_\odot$ and $P=50$ days. The initial $Z$ values for the binary models are $0.008$ and $0.002$.

All models begin with the following chemical composition distributed uniformly throughout the star -- {the metal mass fraction $Z$ with mass fractions of individual metals scaling according to the solar abundances taken from \citet{GS98},} the initial He mass fraction $Y$ in our models is calculated as $Y = Y_{\text{prim}} + (\Delta Y/\Delta Z)\, \times\, Z$ for a given $Z$, where the primordial He abundance, $Y_{\text{prim}} = 0.24$ and He enrichment factor, $\Delta Y/\Delta Z = 2$ \citep{Audouze1987, Pols1998}. The H mass fraction $X$ is then given by $X = 1-Y-Z$. All models are evolved until core-He exhaustion {unless specified otherwise.}

\subsubsection{Mixing processes and mass loss}
The core region of massive stars are unstable to convection due to the enormous amounts of energy produced in the center. Radiative diffusion alone cannot transport all the produced energy and convection sets in. The standard mixing length theory (MLT) from \citet{MLT68} is used in our models with a fixed convective mixing length parameter of $\alpha_\mathrm{MLT} = 1.5$. The same mixing length formalism is also employed in the subsurface convective layers of our models that arise around opacity bumps. The MLT++ routine is disabled, meaning that the temperature gradient predicted by MLT is used as-is, without any reduction in the temperature gradient.

The convective boundary is set by the Ledoux criteria which takes into account the composition gradient when comparing the radiative and adiabatic temperature gradients. MESA adopts a diffusive treatment for semi-convection with the diffusion coefficient taken from \citet{Langer1983}. {For our single-star models,} we use a fixed semi-convective factor of  $\alpha_\mathrm{sc} = 1$ \citep{Yoon2006, Abel2019}. 

Convective boundary mixing is present in the form of exponential overshooting above the core, with $f_\mathrm{ov}$ as an adjustable parameter \citep{Herwig2000}. We use a value of $f_\mathrm{ov}$ of 0.03 {for our single-star models.} If alternate values are used for testing purposes, this is mentioned in the text in the corresponding subsections. 

For the single-star models presented in $\mathrm{Sect.\,}\ref{sec: exp_vs_infl}$, mass loss is completely turned off. This is to focus on the underlying structure of the envelope without interference. For single-star models in $\mathrm{Sect.\,}\ref{sec: discussion}$, we subject them to constant mass-loss rates to analyse its impact on the envelope. 

{For our binary models, the mixing and mass-loss inputs, including both wind mass loss and mass-transfer rates via Roche-lobe overflow, follow the inputs used in \citet{Dutta_Klencki2024}. A key parameter tested in our binary grid is the semi-convection factor, which is varied across $\alpha_\mathrm{sc} = 0.01$, $1$, and $100$. The results for $\alpha_\mathrm{sc} = 1$ are very similar to those for $\alpha_\mathrm{sc} = 0.01$. To simplify the discussion, we present the results of the $\alpha_\mathrm{sc} = 0.01$ and $100$ models in this work. }

{All our models are initially non-rotating.}

\subsubsection{Relevant microphysics}

For the reaction net, we use the \texttt{basic.net} network that includes eight isotopes. The opacities used are from the OPAL Type 2 opacity tables, which takes into account of the changes in the mass fractions of metals (mainly C and O) as the star evolves. For the equation of state (EOS), we use the tables available in MESA that are mainly based on the OPAL EOS tables \citep{Rogers2002} plus a blend of other EOS tables \citep[for further details, see][]{MESA11}.

\subsection{Definitions}

The beginning of core-H burning or H-ZAMS is defined when the center burns 0.01 mass fraction of hydrogen, that is, when $X < X_\mathrm{max} - 0.01$. The end of core-H burning or the Terminal age MS (TAMS) is defined at central $X$ of 0.01.  Similarly, the beginning of core-He burning is defined when the center burns 0.01 mass fraction of He, that is, when $Y < Y_\mathrm{max} - 0.01$. The end of core-He burning is defined at central He mass fraction of 0.01. 

Our focus is on the envelope properties of expanding and inflating models, so it is essential to clearly define what constitutes an envelope. We start by defining the core: the term `core' {in this work} refers to the inner He-rich region of the star. During the MS, the `core' specifically refers to the convective core where He is being produced from H. {Beyond the MS, the `core' is defined as the outermost mass co-ordinate where $X$ is below 0.01 and $Y$ is above 0.1, referring to the pure helium part of the star.} The `envelope' then encompasses everything above the core, including the H-burning shell just above the He core. This distinction is crucial because, as we show, not all parts of the envelope undergo inflation.

\section{Comparing expansion and inflation}
\label{sec: exp_vs_infl}
In this section we systematically differentiate between the expansion of stellar evolution models beyond the MS and the phenomenon called inflation. Both these processes result in a rapid increase of the radius of the star but occur under very different circumstances. 

\subsection{Envelope expansion}
\label{sec: redward_exp}

The dwarf-to-giant configuration switch in stellar models once H is exhausted in the core has been the subject of substantial debate over the past few decades \citep[][]{Hoppner1973, Eggleton1981, Applegate1988, Renzini1992, Sugimoto2000, Faulkner2005, Stancliffe2009}. Many explanations have been proposed to understand the cause of this transition, ranging from gravo-thermal instability of the compact core \citep[for e.g.][]{Iben1993}, strengthening of the core gravitational field \citep{Hoppner1973, Weiss1983}, opacity-driven thermal instability of the envelope in causing a runaway expansion \citep{Renzini1992, Renzini1994} to mean molecular weight gradient \citep{Stancliffe2009}. 

In this section, we refrain from delving into the ultimate cause for the dwarf-to-giant configuration switch. Instead, we analyse the general properties of {massive star envelopes during the rapid redward expansion phase}, which are then compared to the phenomenon of inflation.

Stars spend most of their lives in thermal equilibrium which is just a statement of energy balance inside the star. At each layer inside the star, one can write 
\begin{equation}
\begin{array}{c@{\qquad}c}
\dfrac{\partial L_\mathrm{act}}{\partial m} = \dfrac{\partial L_\mathrm{nuc}}{\partial m} + \epsilon_\mathrm{grav} = \epsilon_\mathrm{nuc} +\epsilon_\mathrm{grav},
\end{array}
\label{eq: energy_balance}
\end{equation}
where $L_\mathrm{act}$ is the actual local luminosity inside the star, $L_\mathrm{nuc}$ is the energy generated from nuclear fusion, $\epsilon_\mathrm{nuc}$ is the specific nuclear energy generation rate (in ergs/g/s) and $\epsilon_\mathrm{grav}$ is called the gravitational source function term\footnote{Although it is termed a `source' function, this quantity can be either negative or positive and thus can be a source or a sink locally.}. All these quantities are defined locally. The $\epsilon_\mathrm{grav}$ term is given by 
\begin{equation}
\begin{array}{c@{\qquad}c}
\epsilon_\mathrm{grav} = -c_\mathrm{p} T \Big[ (1-\nabla_\mathrm{ad}\chi_T)\dfrac{D\mathrm{ln}T}{Dt} - \nabla_\mathrm{ad}\chi_{\rho}\dfrac{D\mathrm{ln}\rho}{Dt}\Big] + \epsilon_\mathrm{grav, X} 
\end{array}
\label{eq: grav_source_func}
\end{equation}
\citep{Kipp1990}, where $c_\mathrm{p}$ is specific heat at constant pressure, $\nabla_\mathrm{ad}$ is the adiabatic temperature gradient, $\chi_T = (\partial \mathrm{ln} P/\partial \mathrm{ln} T)_{\rho}$, \; $\chi_{\rho}=(\partial \mathrm{ln} P/\partial \mathrm{ln} \rho)_{T}$ are EOS exponents, and $\epsilon_\mathrm{grav, X}$ captures the composition dependence of the internal energy.

\begin{figure}
    \includegraphics[width = \columnwidth]{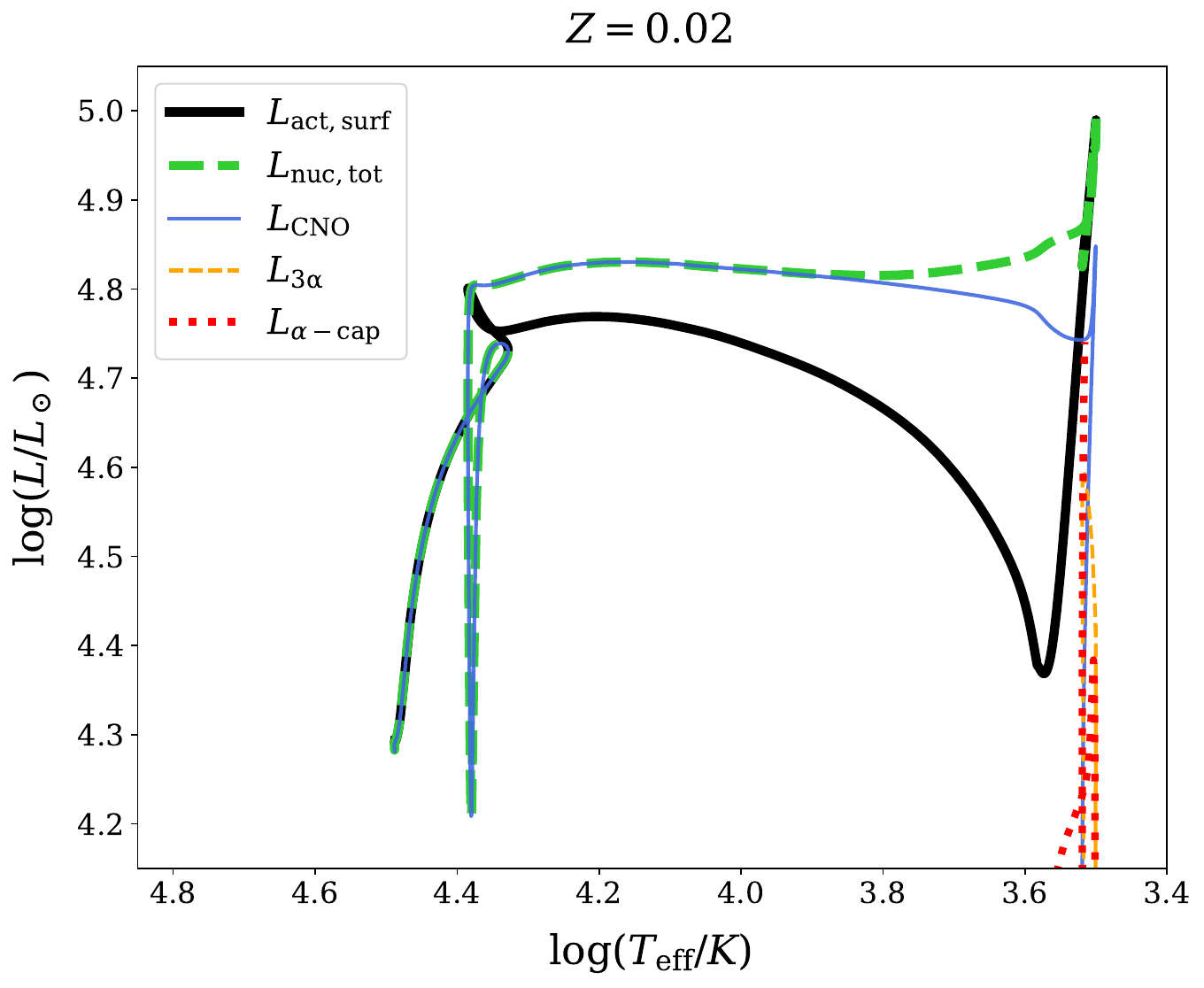}
    \caption{Hertzsprung-Russell diagram track of a \(15\,M_\odot\) model. {The surface and total nuclear luminosity are shown by the black solid and green dashed lines, respectively. The contributions to the nuclear luminosity include the CNO cycle during H burning (blue), the triple-\(\alpha\) reaction (orange dashed), and additional alpha captures (red dotted) like $^{12}\mathrm{C}(\alpha, \gamma)^{16}\mathrm{O}(\alpha, \gamma)^{20}\mathrm{Ne}$ and \(^{14}\mathrm{N}(\alpha, \gamma)^{18}\mathrm{F}(, e^+)^{18}\mathrm{O}\) during core-He burning. After core-H exhaustion, the model rapidly \textit{expands} into an RSG. Notice the increasing difference between actual and nuclear luminosities, especially below \(\log(T_\mathrm{eff}/K) \sim 4\).}
} 
    \label{fig: HRD_15os3}
\end{figure}

This quantity $\epsilon_\mathrm{grav}$ is critical in understanding the evolution of surface $L_\mathrm{act}$ and $T_\mathrm{eff}$  \citep[e.g.][]{Kipp1990, Maeder2009, Farrell2021}. The term $\epsilon_\mathrm{grav}$ can be understood as the excess energy that is not easily dissipated, leading to a local expansion, or the deficit of energy that prompts the layers to contract in response. If the nuclear energy produced in a shell, $\Delta L_\mathrm{nuc} = \epsilon_\mathrm{nuc} \Delta m$ is greater (or lesser) than the energy that is transported outwards, $\Delta L_\mathrm{act}$, a \textit{local expansion (or contraction)} occurs and $\epsilon_\mathrm{grav}$ is negative (or positive). When this $\epsilon_\mathrm{grav}$ term becomes negligible, and $\Delta L_\mathrm{act} \approx \Delta L_\mathrm{nuc}$ in each shell (which also means $ L_\mathrm{act} \approx  L_\mathrm{nuc}$ at each layer), the model is in \textit{thermal balance}.

Consider a H-burning model that is in thermal balance. Due to core-H burning, {H decreases in favour of He and the mean molecular weight in the core increases}, leading to rises in the core temperature and density. Consequently, the nuclear luminosity increases within the core as $\epsilon_\mathrm{nuc} = \epsilon_\mathrm{nuc}(\rho, T, X_i)$, causing a mismatch between the nuclear and actual luminosity. Or in other words, the energy produced in the core is not the same as the energy lost at the surface. The actual luminosity adjusts to this mismatch on a thermal timescale, with the envelope layers locally expanding to accommodate the higher nuclear luminosity. If this local expansion allows for a higher actual luminosity transport outwards, then the model regains thermal balance. 

\begin{figure}
    \includegraphics[width = \columnwidth]{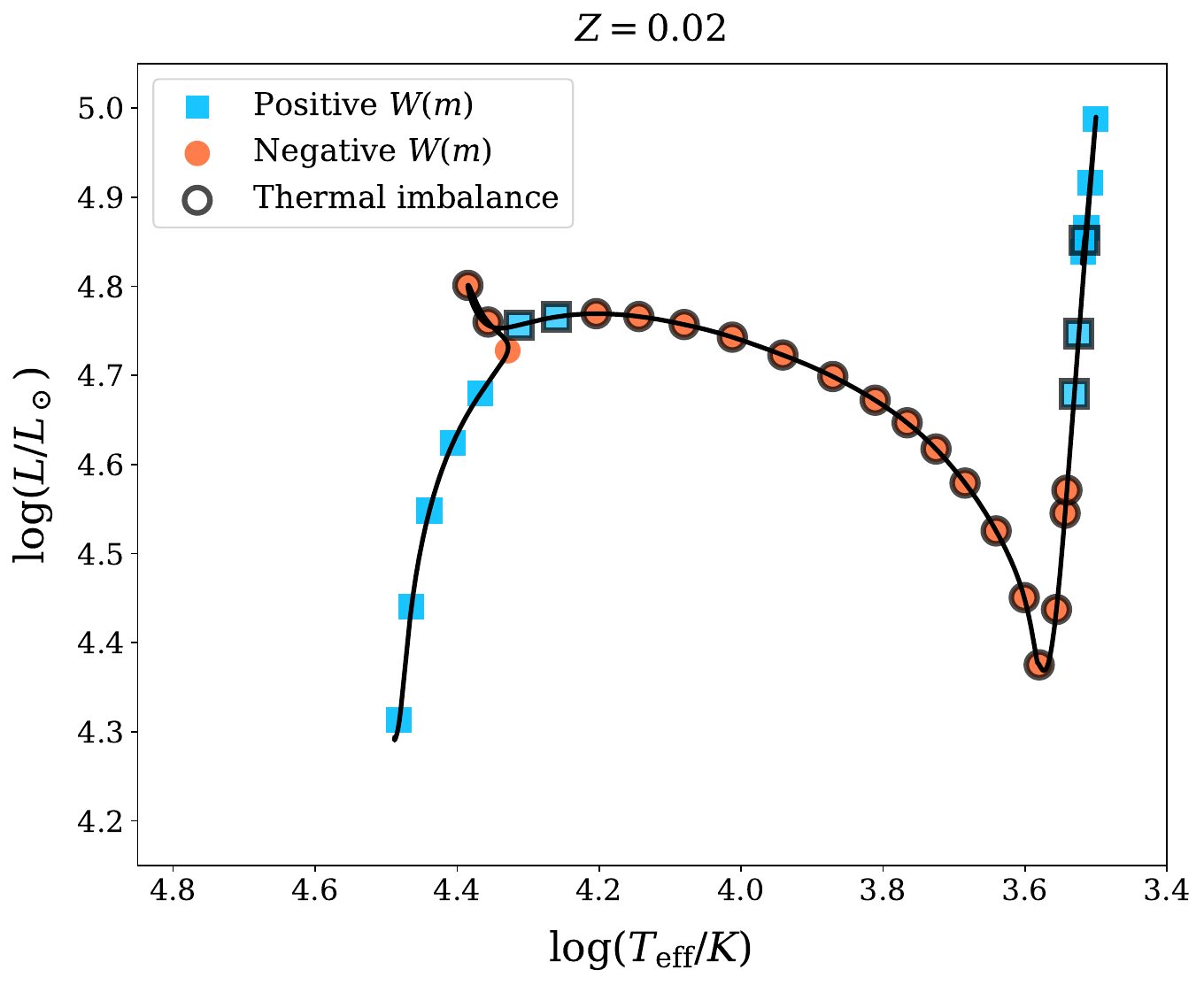}
    \caption{ Hertzsprung-Russell diagram track of the \(15\,M_\odot\) model in $\mathrm{Fig.}\,\ref{fig: HRD_15os3}$, but over-plotted with coloured symbols based on the value of \(W_{m}\) in the envelope. Blue squares if positive \(W_{m}\), reflecting the envelope's ability to regain thermal balance, while red circles if negative \(W_{m}\), indicating an unstable scenario. Symbols are highlighted with a thick black border if the deviation from thermal balance exceeds \(1\%\) between actual and nuclear luminosity.}
 
    \label{fig: HRD_exp_condition}
\end{figure}

\citet{Renzini1992} captures the capability of the envelope to adjust itself to adapt to the changes in the nuclear luminosity by a quantity called $W_{m}$,  which essentially examines the variations in the actual luminosity due to radial adjustments in the envelope (at constant $m$) between successive model configurations. The quantity $W_{m}$ is defined by

\begin{equation}
\begin{array}{c@{\qquad}c}
W_{m} = \dfrac{\delta \mathrm{ln}L_\mathrm{act}}{\delta \mathrm{ln}r} \Bigg]_m   =  \dfrac{\mathrm{ln} L_\mathrm{act, new} - \mathrm{ln} L_\mathrm{act, old}}{\mathrm{ln} r_\mathrm{new} - \mathrm{ln} r_\mathrm{old}} \Bigg]_m 
\end{array}
\label{eq: Renzini_w}
\end{equation}
at each layer $m$, where `new' and `old' correspond to successive configurations. If $W_{m}$ is positive throughout the envelope, then the model can regain thermal balance in the event of any sudden departure from thermal balance, i.e., the thermal balance is stable. However if $W_{m}$ is negative, the envelope adjustment only exacerbates the thermal imbalance problem leading to an unstable situation.

In $\mathrm{Fig.}\,\ref{fig: HRD_15os3}$, we show the Hertzsprung-Russell (HR) diagram evolution track of our $15\,M_\odot$, {showcasing the difference between the actual surface luminosity and the total nuclear luminosity.} Also shown are the various contributions to this nuclear luminosity. {In $\mathrm{Fig.}\,\ref{fig: HRD_exp_condition}$, we indicate the nature of $W_{m}$ at various stages of evolution by comparing each model configuration with the immediate next model step.}  Only envelope layers with temperatures cooler than log($T$/K) $< 7$ are taken into consideration when evaluating $W_{m}$, as it covers all the prominent opacity bumps due to line and continuum transitions. Additionally, we also indicate whether each configuration is in thermal balance or not.

During the MS, the model is in thermal balance and $W_{m}$ is positive indicating a stable thermal balance. The energy produced in the core by nuclear fusion (through the CNO cycle) provides the entire luminosity, that is, $ L_\mathrm{act} \approx  L_\mathrm{nuc}$ at each $r$. As H burns, the nuclear luminosity $L_\mathrm{nuc}$ increases. However, given that $W_{m}$ is positive, the actual luminosity can adapt to this higher nuclear luminosity, that is, the resulting local expansion of the envelope layers tends to reduce the difference between the two luminosities. Gradually the envelope evolves towards larger radii during the MS, with the actual luminosity increasing to match the rise in the nuclear luminosity, {which is essentially a mean molecular weight ($\mu$--)effect}.

\begin{figure}
    \includegraphics[width = \columnwidth]{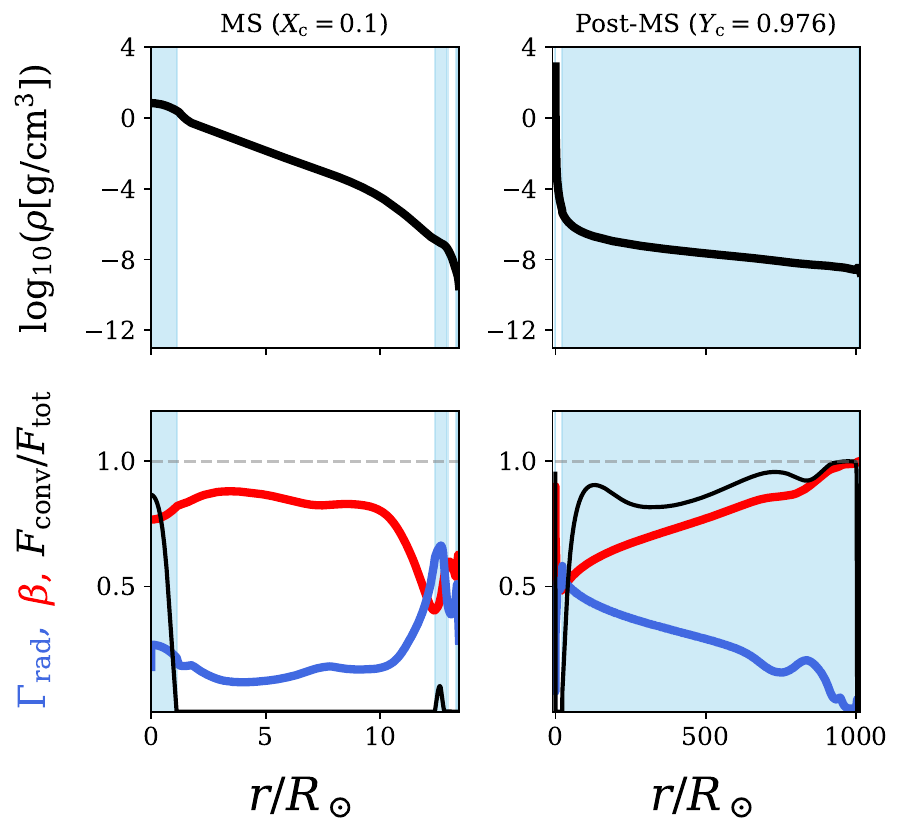}
    \caption{Internal profiles from the \(15\,M_\odot\) model taken near the end of the MS before expansion (left) and at the onset of core-He burning after expansion (right). The profiles show the stratification  of density (top panel), local \(\Gamma_\mathrm{rad}\), gas-to-total pressure ratio \(\beta\), and convection flux fraction (bottom panel). The {blue} shaded region indicates convective regions. The post-expansion model exhibits about nine orders of magnitude density contrast between the He-rich core and the envelope.
}
    \label{fig: expanded_15msun}
\end{figure}

However, deviations from energy balance can occur when the core's nuclear fusion source begins to deplete. This is seen by a rapid decrease in the nuclear luminosity towards the end of the MS in $\mathrm{Fig.}\,\ref{fig: HRD_15os3}$. The ongoing MS expansion is halted and the entire model contracts resulting in a `hook'-like trajectory in the HR diagram. In this phase, the model progressively moves away from thermal balance as seen by the red circles in $\mathrm{Fig.}\,\ref{fig: HRD_exp_condition}$. Not only is there a thermal imbalance, but with each subsequent step the model strays further away from thermal balance, indicating an unstable situation. This is due to the rapid depletion of core H, and the inability of the actual luminosity in the envelope to adjust to this sudden drop in the nuclear luminosity.

The convective nature of massive star cores results in simultaneous depletion of H throughout the entire core.  If by the end of core-H burning, the He core mass is larger than a critical Schönberg–Chandrasekhar limit (of about $M_\mathrm{He,core}/M_\mathrm{tot}\,\sim0.1$) \citep{Schonberg1942}, then the core cannot remain stable and rapidly contracts which brings the H-rich layers just above the He core to temperatures hot enough to burn H in a shell around the core. This is seen by the rapid recovery in the nuclear luminosity following H exhaustion. 

\begin{figure}
    \includegraphics[width = \columnwidth]{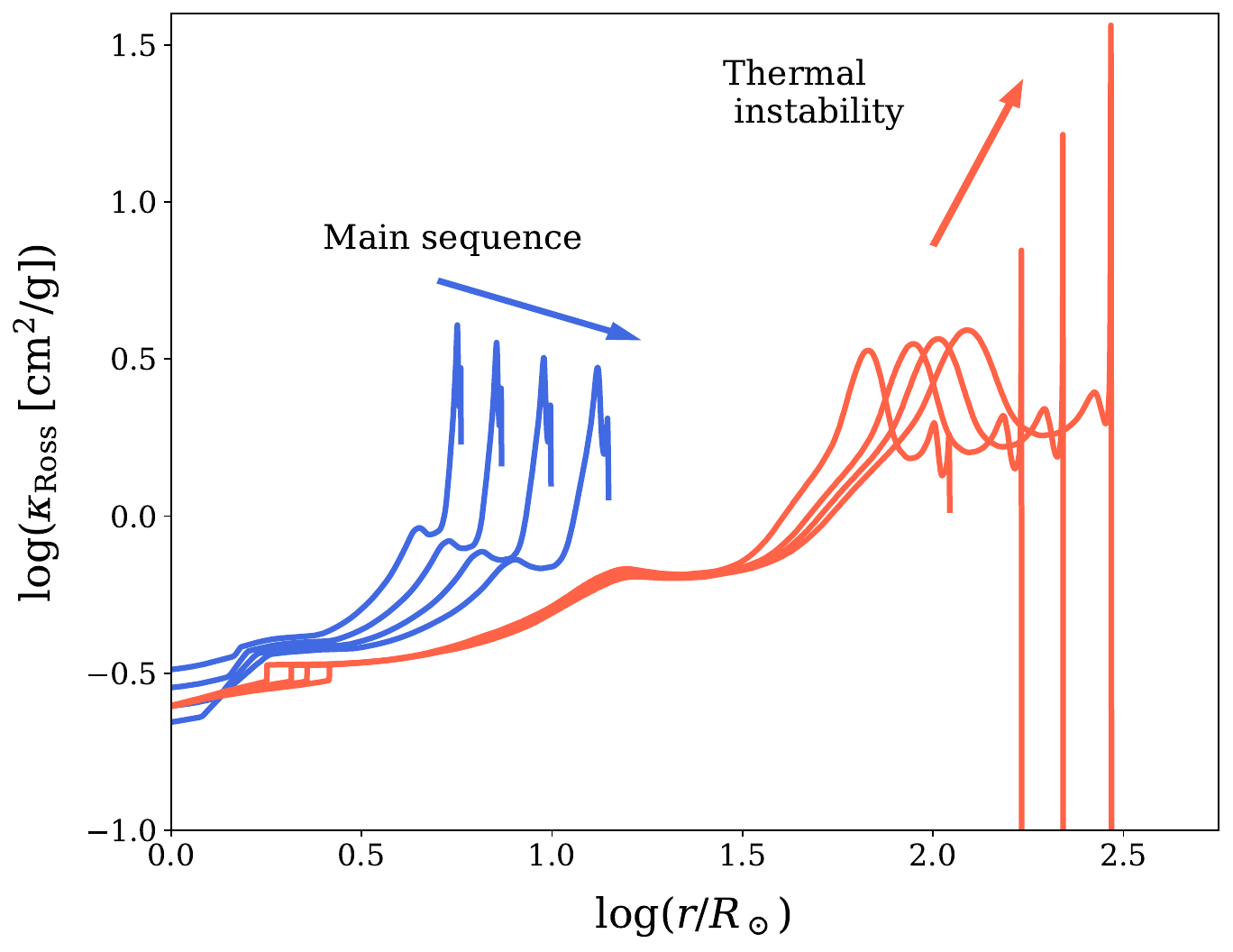}
    \caption{Radial stratification of the Rosseland mean opacity during the MS (blue) and the runaway thermal instability phase below log($T_\mathrm{eff}/K$) $\sim4$ (red).
}
    \label{fig: 15msun_opacity_strat}
\end{figure}

The stable thermal balance phase during the MS stands in stark contrast to the evolution following the complete depletion of H in the core. The core continues to rapidly contract, increasing the temperature and density in the H-burning shell. The nuclear luminosity increases over the actual luminosity, that is, the gravitational source term $\epsilon_\mathrm{grav}$ in the envelope is \textit{negative and non-negligible}, resulting in a local expansion in the envelope layers. Initially this local expansion tends to bring the $15\,M_\odot$ model to thermal balance as evident by the positive $W_{m}$ in $\mathrm{Fig.}\,\ref{fig: HRD_exp_condition}$ immediately after the hook. However, the overall model is far from thermal balance (cf. $\mathrm{Fig.}\,\ref{fig: HRD_15os3}$) and the  model continues to expand.

The negative $W_{m}$ values quickly reappear however, as this expansion leads to an increase in the opacity (cf. $\mathrm{Fig.}\,\ref{fig: 15msun_opacity_strat}$) which tends to reduce the actual luminosity. The nuclear and actual luminosity begins to deviate from each other, giving rise once again to an unstable, thermal imbalance situation. This is particularly apparent beyond log$(T_\mathrm{eff}/K) \sim4$ in $\mathrm{Fig.}\,\ref{fig: HRD_15os3}$, coinciding with the onset of a new opacity bump {--- the H-bump ---} which becomes stronger as the model evolves towards cooler temperatures. Regardless of whether such thermally unstable envelopes are a correlation or a causation for the dwarf-to-giant transition, the model rapidly moves away from thermal balance from this point onwards. The quantity $W_{m}$ eventually becomes positive when a deep convective zone develops sweeping inwards. The almost fully convective model settles on the Hayashi line \citep{Hayashi1961, Kipp1990} and increases in luminosity and radius while keeping the temperature nearly constant. 

{We would like to clarify two points here. First,} regarding the nomenclature in this work, we refer to the evolution immediately following the hook, until the model restores thermal balance (as an RSG in this case), as the \textit{expansion} (or \textit{expanding}) phase. Throughout this phase, the model experiences thermal imbalance, characterized by a non-negligible negative $\epsilon_\mathrm{grav}$ in the envelope. Part of this expansion phase involves a runaway \textit{thermal instability}, during which the model rapidly moves away from thermal balance.

{Second, the loss of thermal balance during the redward expansion phase  described above applies to intermediate-mass and massive star models, where this process occurs on a thermal timescale. In contrast, low-mass stars ($\lesssim1.1\, M_\odot$) can achieve red giant dimensions on a nuclear timescale without losing thermal balance \citep[for further discussions and debates on this, see, e.g.][]{Faulkner2005, Miller2022, Renzini2023}. In this work, we focus on the dwarf-to-giant transition in massive-star models after core-H exhaustion, where the aforementioned properties hold true. These can later be directly compared to another mechanism that is relevant for massive stars -- inflation.}

Thermal balance is eventually regained in our $15\,M_\odot$ model as the He-rich core becomes hot enough to burn He through the triple-$\alpha$ process. The density stratification during this stable phase has a clear imprint of the previously occurred expansion phase -- a very dense core above which the density drops steeply and a low-density, but massive envelope on top of it. We refer to such envelopes as \textit{expanded} envelopes. The term `expanded' here simply indicates that the envelope has experienced a phase of expansion marked by thermal imbalance prior to stabilizing in a nuclear burning phase; it does not imply that the model is currently out of thermal balance or undergoing expansion.

\begin{figure}
    \includegraphics[width = \columnwidth]{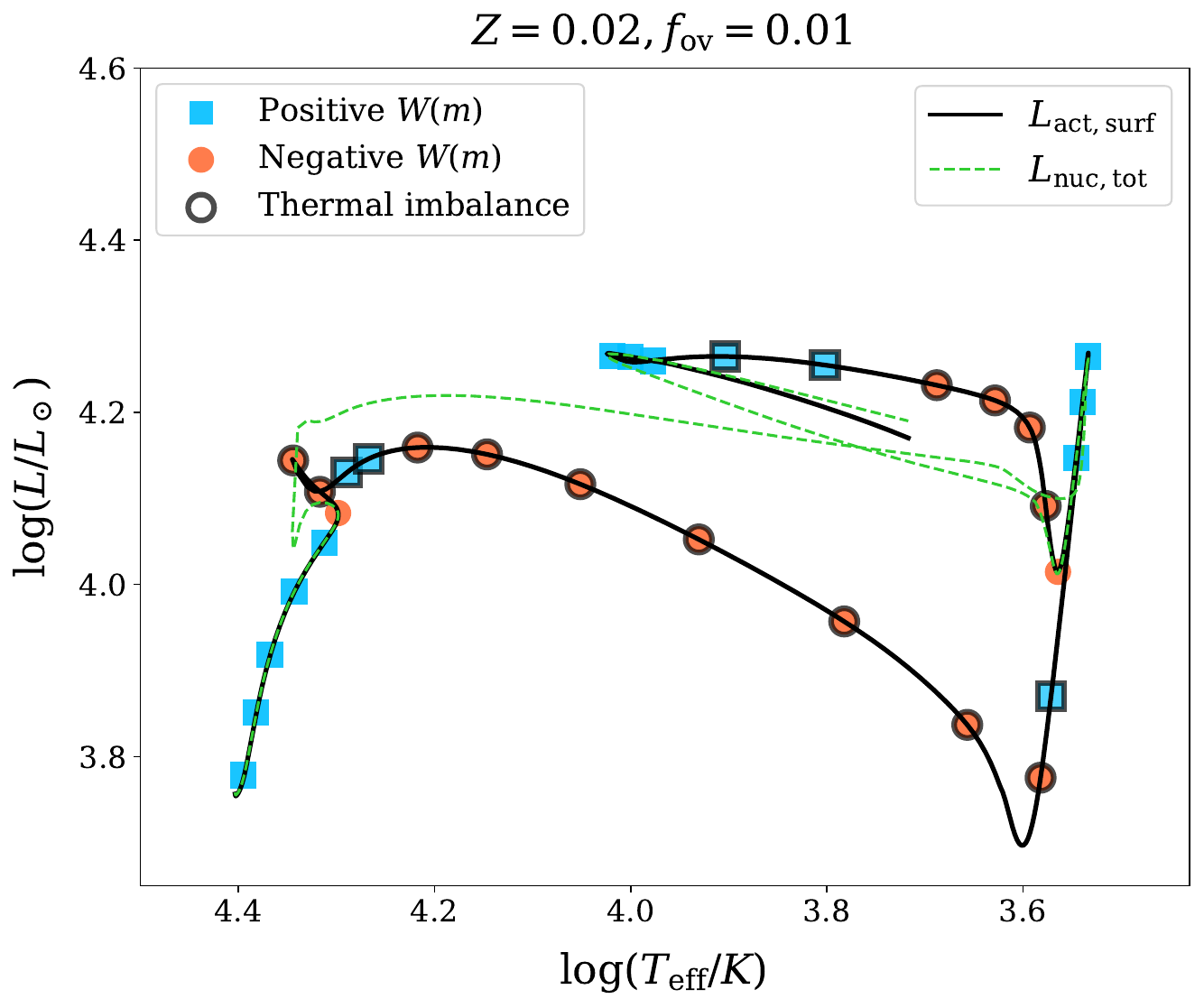}
    \caption{{Hertzsprung-Russell  diagram of a \(10\,M_\odot\) model with low MS core overshooting, illustrating a blue loop contraction episode. The lines and coloured symbols have the same meaning as in $\mathrm{Figs.\,}\ref{fig: HRD_15os3}\;\mathrm{and}\;\ref{fig: HRD_exp_condition}$. An additional term \(W_{\mathrm{nuc},m}\) is subtracted to assess stability during the blue loop phase. Roughly three-quarters of the way through core-He burning as an RSG, the model undergoes a blue loop, initially moving away from thermal balance before regaining stability and continuing as a BSG.}
} 
    \label{fig: blue_loop}
\end{figure}

In $\mathrm{Fig.}\,\ref{fig: expanded_15msun}$, we show the density stratification of the $15\,M_\odot$ model close to end of the MS ($X_\mathrm{c} = 0.1$) before the expansion phase, and at the onset of core-He burning ($X_\mathrm{c} = 0, Y_\mathrm{c} = 0.976$) after the expansion phase has taken place. The density in the core of the expanded model is approximately two orders of magnitude higher compared to the core during the MS phase. Just above the core, the density drops by approximately 9 orders of magnitude. The envelope prior to expansion weighed nearly $10\,M_\odot$ in approximately ten solar radii. After expansion, the envelope still weighs roughly $10\,M_\odot$, but now the mass is distributed across a thousand solar radii. This is what we mean by `low-density, but massive' envelope. The radial adjustment that occurs during the expansion phase is across the entire envelope. {$\mathrm{Fig.}\,\ref{fig: expanded_15msun}$ also shows the internal variation of the radiative Eddington parameter (cf. $\mathrm{Eq.\,}\ref{eq: Eddington_parameter}$).}

The core-He-burning phase in our $15\,M_\odot$ model is started and finished as an RSG. However, this is not always the case. {Whether models stabilize as blue or red supergiants is highly sensitive to the internal chemical profile, which is influenced by various mixing processes within the model.} Models with a relatively smaller core size (either due to low convective boundary mixing or low mass loss) or a steeper H/He composition gradient -- {for example from strong semi-convective mixing} --  can stabilize as blue supergiants (BSGs). In such scenarios, one might observe transitions from an RSG to a BSG configuration during core-He burning, leading to the formation of so-called blue loops \citep{Hayashi1962, Schlesinger1977, Stothers1979, Langer1983, LangerMaeder1995, Maeder2009, Abel2019, Higgins2020, Szcsi2022, Farrell2021}. Such loops are typically weaker or nonexistent {in models with strong mass loss }which typically occurs at higher metallicity and higher initial masses. 

{In $\mathrm{Fig.}\,\ref{fig: blue_loop}$, we show the HR diagram track of a $10\,M_\odot$ model with low core overshooting of $f_\mathrm{ov} = 0.01$. The model undergoes an expansion phase after core-H exhaustion, settling into core-He burning in an RSG configuration. During core-He burning, the model slowly descends the Hayashi track on a nuclear timescale, doing so stably. However, about three-quarters of the way through core-He burning ($Y_\mathrm{c}\sim 0.24$), a runaway thermal instability is triggered once again by the H-bump.} This switch from an RSG to a BSG configuration mirrors the expansion phase described above but in reverse, in the sense that a local contraction in the envelope results in a reduction in the opacity \citep[see][ for more details regarding stability criteria during blue loop episodes]{Renzini1992}.  {The model experiences a blue loop \textit{contraction} episode, rapidly moving away from thermal balance with a non-negligible positive $\epsilon_\mathrm{grav}$ in the envelope. Eventually, the model regains thermal balance in a BSG configuration and resumes core-He burning in a stable fashion.}  Throughout this text, we reserve the words `expansion' and `contraction' for the above described phenomenon. 

The variations in local luminosity within the envelope, and hence the $W_{m}$ parameter, are largely influenced by the envelope opacity\footnote{This is not the only dependence though as $W_{m}$ depends separately on the response of density and temperature to changes in the local radius.}. The stratification of the local Rosseland mean opacity $\kappa_\mathrm{Ross}$ (in $\mathrm{cm}^2/\mathrm{g}$), both during the MS and expansion phase, is shown in $\mathrm{Fig.}\,\ref{fig: 15msun_opacity_strat}$. During the MS, the opacity reduces as the model gradually evolves to higher radii enabling the actual luminosity to adapt to the increase in nuclear luminosity. {On the other hand, during the expansion phase, the opacity increases with local expansion leading to an unstable scenario. }

\vspace{0.5cm}
\noindent
We briefly summarise the properties of expanding (and contracting) envelopes during the evolution of {intermediate-mass and massive} stellar models.
\begin{itemize}
    \item When there is a discrepancy between the actual and nuclear luminosity in any shell, local expansion or contraction takes place. If the adjustment tends to reduce the discrepancy, the model stays in (or moves towards) thermal balance.
    \item Conversely, if the adjustment causes the model to stray further away from thermal balance, then the model precipitously moves away from thermal balance resulting in a runaway instability scenario. Thermal balance is then restored if a new source of nuclear burning emerges (such as during the end of the overall contraction phase) or if a deep convective layer initiates (for example, towards the end of the expansion phase).
    \item The expansion (or contraction) phase {in intermediate-mass and massive star models} is characterized by thermal imbalance with a non-negligible negative (or positive) $\epsilon_\mathrm{grav}$ in the envelope. A part or whole of the expansion (or contraction) phase can be spent in a runaway thermal instability phase.
    \item A large fraction of the envelope participates in the radial redistribution of mass during the expansion phase. Models after the expansion phase, or simply expanded models, can have tens of solar masses residing in thousands of solar radii, that is, the envelope is very low density, yet massive.
\end{itemize}

\subsection{Inflation and how it is different}
\label{sec:infl}

\begin{figure*}
    \includegraphics[width = \textwidth]{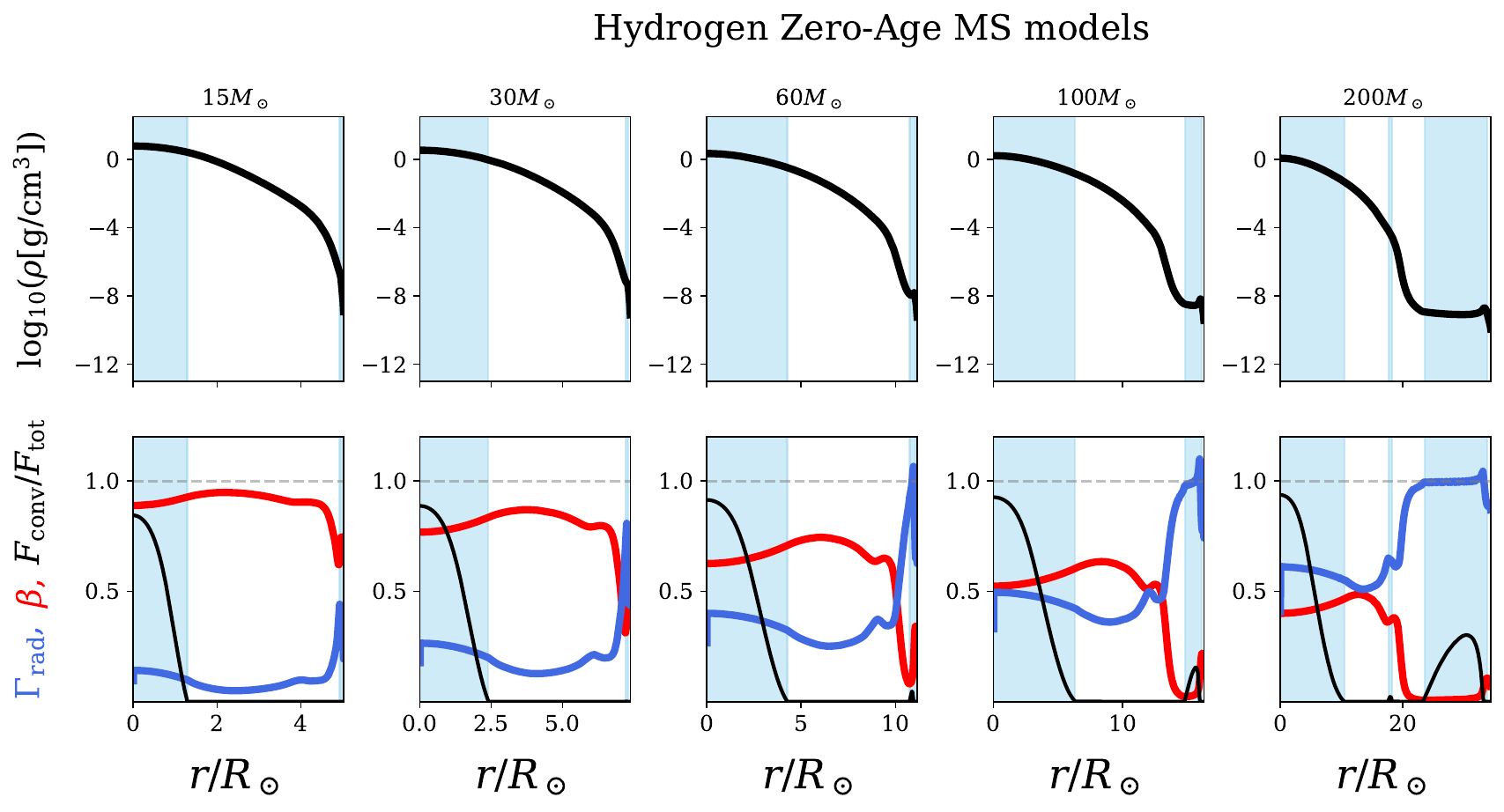}
    \caption{{Stratification of density (top row), local $\Gamma_\mathrm{rad}$, gas-to-total pressure ratio $\beta,$ and convective flux fraction (bottom row) for models on the H-ZAMS.} The five columns represent initial masses from 15 to \(200\,M_\odot\). The {blue} shading indicates convective areas. {The plot illustrates the envelope response as \(\Gamma_\mathrm{rad}\) nears the Eddington limit, leading to inflated morphology for the highest masses. All models are in hydrostatic and thermal balance.}
}
    \label{fig: inflation_prof_ZAMS}
\end{figure*}

Inflation refers to the phenomenon where stellar models develop extended, low mass, low density layers on top of a dense base when approaching their so-called Eddington limit \citep{Ishi1999, Petrovic2006, Graf2012, Sanyal2015}. 


The existence of inflated layers in Nature is debatable, {with evolutionary codes employing different methods to handle them, either by allowing the models to inflate or by implementing strong mass loss and/or efficient energy transport \citep{Kohler2015, Ekstrom2012, MESA13, Agrawal2022}}. Strong mass loss could potentially destroy such low density, inflated layers. Existing simulations show large turbulent velocities near sub-photopsheric layers leading to shocks and density fluctuations that can enable photons to preferentially escape through low-density regions \citep{Shaviv1998, Owocki2004, Jiang2015, Debnath2024}. These effects could potentially reduce the effect of envelope inflation. In this work, we neither argue for or against the existence of inflated layers. We are merely interested in properties of such layers as predicted by 1D hydrostatic structure and evolution models using standard MLT.

The Eddington limit can be defined locally when the radiative acceleration due to transfer of momentum from photons to ions equals the force of gravity. \citet{Sanyal2015} define an appropriate local radiative Eddington parameter taking into account only the radiative luminosity. We use the same definition here:
\begin{equation}
\begin{array}{c@{\qquad}c}
\Gamma_\mathrm{rad} = \dfrac{a_\mathrm{rad}}{g} = \dfrac{\kappa_\mathrm{Ross}L_\mathrm{rad}}{4\pi Gc m} = \dfrac{L_\mathrm{rad}}{L_\mathrm{Edd}},
\end{array}
\label{eq: Eddington_parameter}
\end{equation}
where $\kappa_\mathrm{Ross}$ is the Rosseland mean opacity in $\mathrm{cm}^2/\mathrm{g}$ units, $L_\mathrm{rad} = L_\mathrm{act}-L_\mathrm{conv}$ is the energy transported by radiation alone and $L_\mathrm{Edd}$ is the Eddington luminosity defined as  
\begin{equation}
\begin{array}{c@{\qquad}c}
L_\mathrm{Edd} = \dfrac{4\pi Gcm}{\kappa_\mathrm{Ross}}.
\end{array}
\label{eq: Eddington_lum}
\end{equation}
The second equality in $\mathrm{Eq.\,}\ref{eq: Eddington_parameter}$ (that includes the Rosseland mean opacity) only holds within the interior of the star under conditions of sufficiently high optical depths, where radiative diffusion is applicable. All quantities in the above two equations are locally defined.

This distinction between the actual and radiative luminosity, and the corresponding  Eddington parameters ($\Gamma$ vs. $\Gamma_\mathrm{rad}$) is important. For example, in the core, the energy produced by nuclear burning (which will be equal to the actual luminosity for thermal balance) is well above the local Eddington luminosity. However the convective instability is triggered first, and convection carries most of the flux ($\sim90\%$ of the total flux) before the radiative Eddington limit is breached. The two $\Gamma$'s are equal in radiative zones and cooler envelope layers where convection can be inefficient. Unlike convection in the dense core, convection in cool envelope layers around strong opacity bumps are highly inefficient due to lower densities. The majority of flux is carried by radiation despite the layers being convectively unstable, that is, $L_\mathrm{act} \approx L_\mathrm{rad}$. 

The hydrostatic equation (or force balance) connects the relevant forces at each layer inside the model:
\begin{equation}
\begin{array}{c@{\qquad}c}
\dfrac{\partial P_\mathrm{tot}}{\partial m} = \dfrac{\partial P_\mathrm{gas}}{\partial m} + \dfrac{\partial P_\mathrm{rad}}{\partial m} = \dfrac{\partial P_\mathrm{gas}}{\partial m} -\dfrac{a_\mathrm{rad}}{4\pi r^2} = -\dfrac{g}{4\pi r^2},\end{array}
\label{eq: force_balance}
\end{equation}
where $P_\mathrm{gas}(\rho, T)$ and $P_\mathrm{rad}(T)$ are the gas and radiation pressure respectively. Their gradients given by $\partial P_\mathrm{gas}/\partial m$ and $\partial P_\mathrm{rad}/\partial m$ balance the gravitational force. Multiplying the entire equation by $1/\partial P_\mathrm{rad}$ and rearranging the terms, the hydrostatic equation becomes
\begin{equation}
\begin{array}{c@{\qquad}c}
\beta\dfrac{\partial \mathrm{ln} P_\mathrm{gas}}{\partial \mathrm{ln} P_\mathrm{tot}} + \Gamma_\mathrm{rad} = 1,
\end{array}
\label{eq: force_balance_new_eq}
\end{equation}
where $\beta = P_\mathrm{gas}/P_\mathrm{tot}$ is the gas-to-total pressure ratio. Moving outwards from the center, $\partial \mathrm{ln} P_\mathrm{tot}$ is negative as gravity always points inwards. Therefore, depending on the value of $\Gamma_\mathrm{rad}$ locally, the gas pressure gradient term changes and can even become positive, that is, a local super-Eddington layer is accompanied by gas pressure increasing outwards. Since $P_\mathrm{gas} = P_\mathrm{gas}(\rho, T)$ and considering the usual decrease in temperature outwards, a gas pressure inversion implies a corresponding density inversion\footnote{The opposite is not true, i.e. a density inversion might not mean gas pressure inversion. See \citet{MESA13} or \citet{Sanyal2015} for exact condition for density and gas pressure inversions.}. 


The density stratification of five models at the H-ZAMS: 15, 30, 60, 100 and $200\,M_\odot$ (from left to right in order) is shown in $\mathrm{Fig.}\,\ref{fig: inflation_prof_ZAMS}$. The less massive 15 and $30\,M_\odot$ models show a typical decline in the density, but as the initial mass increases, we see the outer layers develop an `elephant trunk-like' inflated morphology. The well-developed inflated morphology of the $200\,M_\odot$ model spans nearly half of its radius while comprising only $10^{-5}\,M_\odot$ of mass. Based on the stratification of the $200\,M_\odot$ model, there are two distinct regions. Above roughly $20\,R_\odot$, the density (and the gas pressure) drops significantly and the profile stagnates with radius and even displays an inversion in this case. These layers are said to be inflated and sit on top of a non-inflated base\footnote{The region we call `base' has been termed `non-inflated core' or just `core' in the literature before. We refrain from terming it `core' as it can be easily confused with the convective core. So we stick with `base' here.}. The appearance of such inflated layers causes the H-ZAMS to bend towards cooler temperatures. At Galactic $Z$, the H-ZAMS bending occurs for initial masses above $100\,M_\odot$ {\citep[for eg., see $\mathrm{Fig.}\,2$ in][]{Sanyal2017}}. Such bending is also seen on the He-ZAMS which is the locus of pure He models, albeit at lower masses \citep{Graf2012}.

\begin{figure}
    \includegraphics[width = \columnwidth]{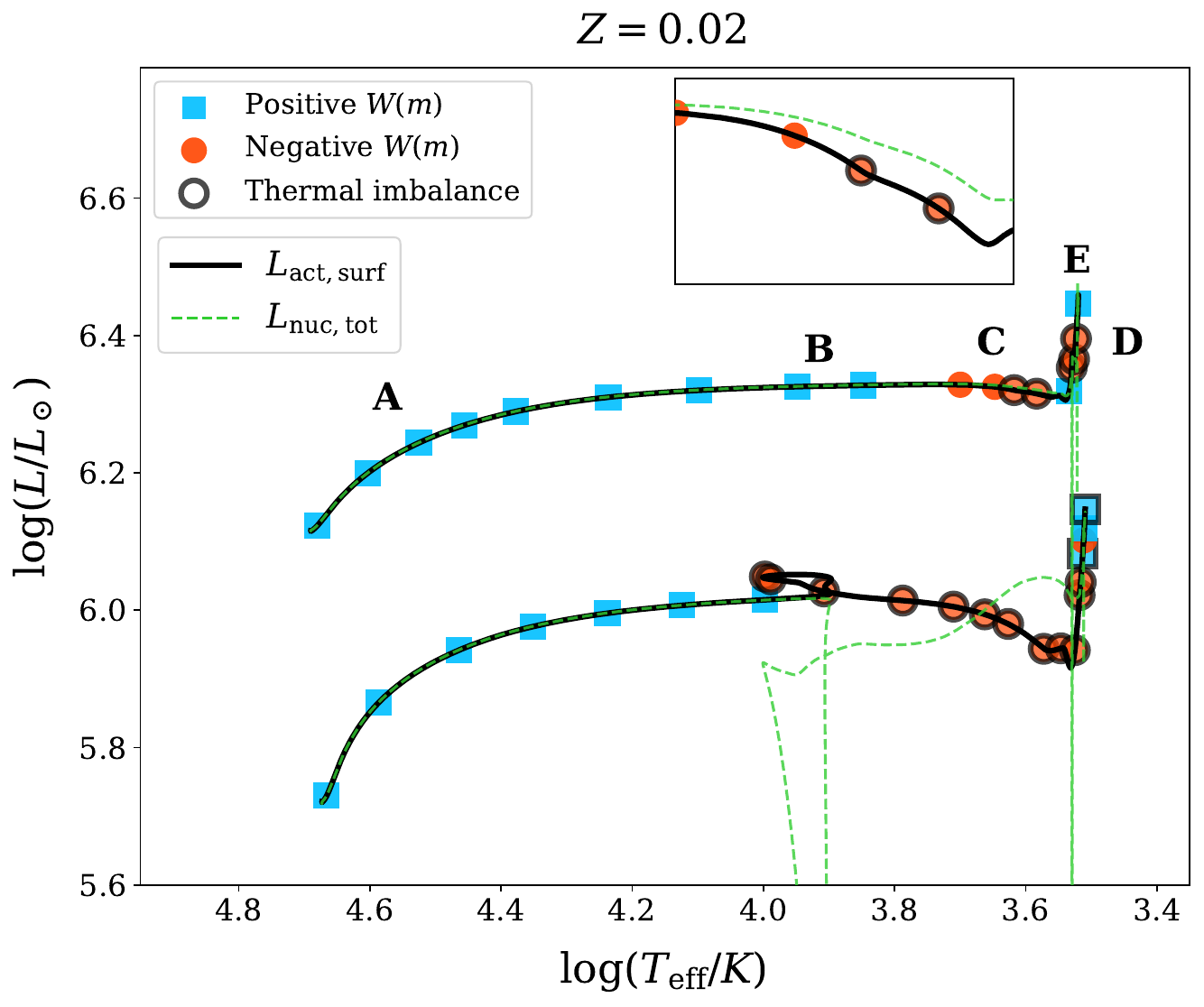}
    \caption{{Hertzsprung-Russell  diagram tracks of 60 and \(100\,M_\odot\) models until core-He exhaustion. The lines and coloured symbols have the same meaning as in $\mathrm{Figs.\,}\ref{fig: HRD_15os3}\;\mathrm{and}\;\ref{fig: HRD_exp_condition}$. The \(100\,M_\odot\) model inflates on the MS until it reaches the H-bump at \(\log(T_\mathrm{eff}/K) \sim 4\), after which it loses thermal stability and \textit{expands} across the H-bump. The growing difference between the surface and nuclear luminosity is magnified in the inset plot.}
} 
    \label{fig: HRD_exp_condition_60_100}
\end{figure}

So why do the layers inflate? As the initial mass increases, so does the luminosity once the model settles on hydrostatic and thermal equilibrium. However, the increase in luminosity is not linear with mass: $L_\mathrm{act} \propto M_\mathrm{init}^\mathrm{x}$ where $\mathrm{x} > 1$. Using a simple Eddington model, the power $\mathrm{x}$ is 3 for low masses and approaches unity as the mass becomes infinite. {In $\mathrm{Fig.}\,\ref{fig: inflation_prof_ZAMS}$, we also show the internal variation of the local $\Gamma_\mathrm{rad}$, the gas-to-total pressure ratio $\beta$, and the convective-to-total flux ratio $f$. As the luminosity-to-mass ratio increases with initial mass on the H-ZAMS, these models can be used to examine how envelope layers respond to an increase in $\Gamma_\mathrm{rad}$ when the layers are near their radiative Eddington limit, compared to those farther from it.}

When $\Gamma_\mathrm{rad}$ is sufficiently low, the envelope morphology is insensitive to changes in the radiative flux. However, as the radiative Eddington limit is approached, the envelope morphology becomes highly sensitive to even small changes in the radiative flux. This is because as the radiative flux increases in inefficient convective zones, the layers adjust themselves to keep the  $\Gamma_\mathrm{rad}$ close to unity (see bottom row of the 60, 100, $200\,M_\odot$ models) and they do so while maintaining hydrostatic and thermal balance. This process, in which layers adjust to lower densities to effectively reduce opacity and prevent a super-Eddington situation, is termed \textit{inflation}. At the end of an inflation phase, the envelope layers that have undergone radial adjustment are termed \textit{inflated} layers, and the entire model, is referred to as \textit{inflated}. Not all layers above the core participate in the inflation process, that is, only those envelope layers where $\Gamma_\mathrm{rad}$ is close to unity and $\beta$ close to zero undergo radial adjustment.

As the density reduces, so does the gas pressure and the $\beta$ parameter. The $\beta$ parameter can never be zero though. In $\mathrm{Eq.\,}\ref{eq: force_balance_new_eq}$, as the radiative Eddington limit is reached, the gas pressure stratification flattens, which means it is the slope that becomes zero as the radiative $\Gamma$ equals unity. A super-Eddington layer is allowed from the 1D structure solution as long as it is accompanied by a gas pressure inversion: $\partial \mathrm{ln} P_\mathrm{gas}$ becomes positive. The density and gas pressure stratification in the inflated layers closely mirror the topology of the major opacity bump around which inflation occurs, as in the local minima in the gas pressure and density closely aligns with the local maxima in the opacity \citep[see also $\mathrm{Fig.}\,5$ from ][]{Graf2012}.

\begin{figure*}
    \includegraphics[width = \textwidth]{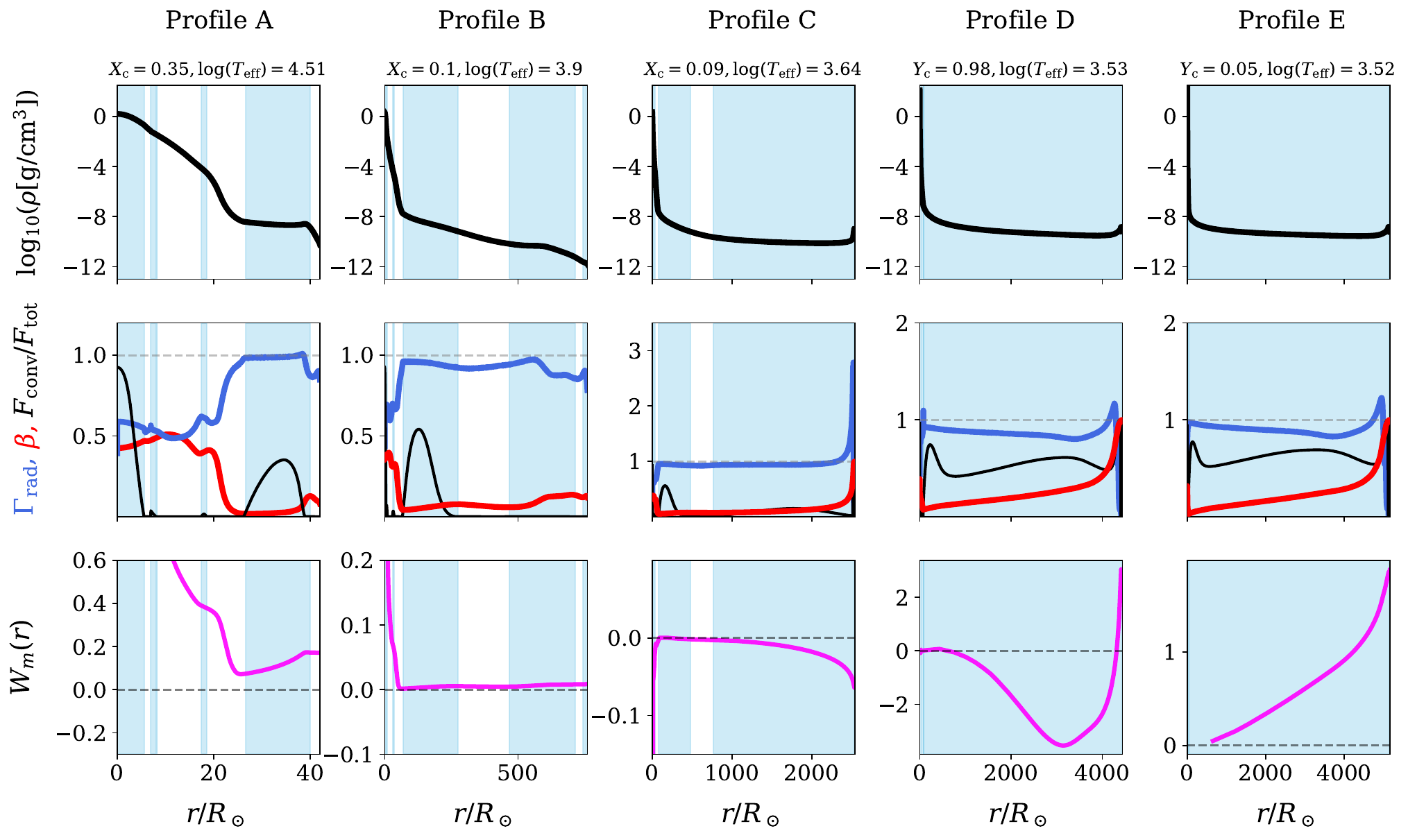}
    \caption{{Stratification of density (top row), local $\Gamma_\mathrm{rad}$, gas-to-total pressure ratio $\beta$, convective flux fraction (middle row), and $W_{m}$ (bottom row)} for five profiles (A-E) taken at important phases of inflation and expansion of the $100\,M_\odot$ model marked in $\mathrm{Fig.}\,\ref{fig: HRD_exp_condition_60_100}$.  {The blue shaded region indicates convectively unstable layers. }} 
    \label{fig: infl_exp_profile_subplots}
\end{figure*}

On the H-ZAMS, models with initial masses above $\sim60\,M_\odot$ show an inflated morphology. So what happens to these layers when the models evolve off the H-ZAMS? {In $\mathrm{Fig.}\,\ref{fig: HRD_exp_condition_60_100}$, we show the HR diagram track of our 60 and $100\,M_\odot$ model.} As before, the models are in thermal balance and $W_{m}$ is positive during the MS, that is, as H is burnt in the core and $L_\mathrm{nuc}$ increases, the envelope layers locally expand to lower opacities. The actual luminosity increases to match the higher nuclear luminosity and the model regains thermal balance {(cf. $\mu$--effect from $\mathrm{Sect.}\, \ref{sec: redward_exp}$)}. However, there are key differences between the MS phase of the models shown here and the $15\,M_\odot$ model from the previous section. First, the increase in absolute luminosity during the MS of the $15\,M_\odot$ is small compared to the  60 and $100\,M_\odot$ models. For example, the absolute increase in the luminosity during the MS, $\Delta L_\mathrm{MS}$, for the 15, 60 and $100\,M_\odot$ models are $3.47 \times 10^{4}, 5.17 \times 10^5$ and $8.97 \times 10^{5}\,L_\odot$ respectively. This is because the core-to-total mass ratio increases with initial mass. A much larger fraction of the star is converted to He leading to a higher $\mu_\mathrm{avg, star}$ and consequently a higher $L_\mathrm{act}$ by the end of the MS. 

Second, the envelope layers of the 60 and $100\,M_\odot$ are already close to their radiative Eddington limit on the H-ZAMS. During the MS, the radiative flux through the inefficient convective zones steadily increases. While this is true for the $15\,M_\odot$ model as well, the proximity of the 60 and $100\,M_\odot$ model to the radiative Eddington limit makes the envelope morphology highly sensitive to the absolute value of $1-\Gamma_\mathrm{rad}$. As the radiative flux steadily increases, the model inflates while attempting to keep the $\Gamma_\mathrm{rad}$ in check, resulting in a large increase in the total radius of the 60 and $100\,M_\odot$ models during their MS phase. For example, the $100\,M_\odot$ inflates to almost two thousand solar radii. {Conversely this also means that models with initial mass lower than $\sim 60\,M_\odot$ can inflate during late MS, for example, \citet{Sanyal2015} find inflated layers in their $M_\mathrm{init}\sim40\,M_\odot$ models during late MS. }

One final point to consider is the role played by processes that control the increase in $L_\mathrm{rad}$ during the MS. Since the radiative Eddington parameter is directly proportional to $L_\mathrm{rad}$, {processes that can potentially increase core size during the MS} can indirectly cause inflation. For example, a large amount of core overshooting during the MS {brings in more fuel for burning. The increase in $L_\mathrm{rad}$ during the MS phase is higher, pushing the $\Gamma_\mathrm{rad}$ parameter closer to its limit. In this scenario, inflated morphology can occur even for initial masses lower than $40\,M_\odot$ found by \citet{Sanyal2015}.} Conversely, a small amount of core overshooting can limit inflation by limiting the increase in $\Gamma_\mathrm{rad}$ during the MS. 

{Yet another process that can increase core size is rotation due to the mixing from rotationally induced instabilities \citep[see][for a review]{MM2000_R, Heger2000}. For low enough rotation, the models end up cooler at the TAMS \citep{Brott2011}, having qualitatively similar effects to higher core overshooting. In the opposite extreme, highly efficient rotational mixing can enable chemically homogeneous evolution, and the model evolves toward the He-ZAMS \citep{Maeder1987, Langer1992}. An inflated morphology may still develop depending on the final mass the model reaches on the He-ZAMS and the amount of H envelope left. A quantitative investigation into the effects of rotation on inflated envelopes is beyond the scope of this work, but it remains a relatively unexplored problem that could be addressed in a future study. }

\vspace{0.5cm}
\noindent
We briefly summarise the properties of an inflating envelope:
\begin{itemize}
    \item  Inflation refers to the phenomenon where stellar models develop very low density envelope layers when approaching their local radiative Eddington limit. The layers inflate to effectively reduce the opacity to try avoid a super-Eddington condition inside the model. 
    \item Inflation can already occur on the MS, that is, the inflated layers are in hydrostatic and thermal balance. 
    \item Not all layers above the core participate in the process of inflation, but only those where the radiative Eddington parameter $\Gamma_\mathrm{rad}$ is very close to unity and the gas-to-total pressure ratio $\beta$ is close to zero. 
    \item The density and gas pressure gradient flattens as the radiative Eddington limit is approached. If it crosses unity, such layers will be accompanied by a positive gas pressure gradient.
\end{itemize}

\begin{figure*}
    \includegraphics[width = \textwidth]{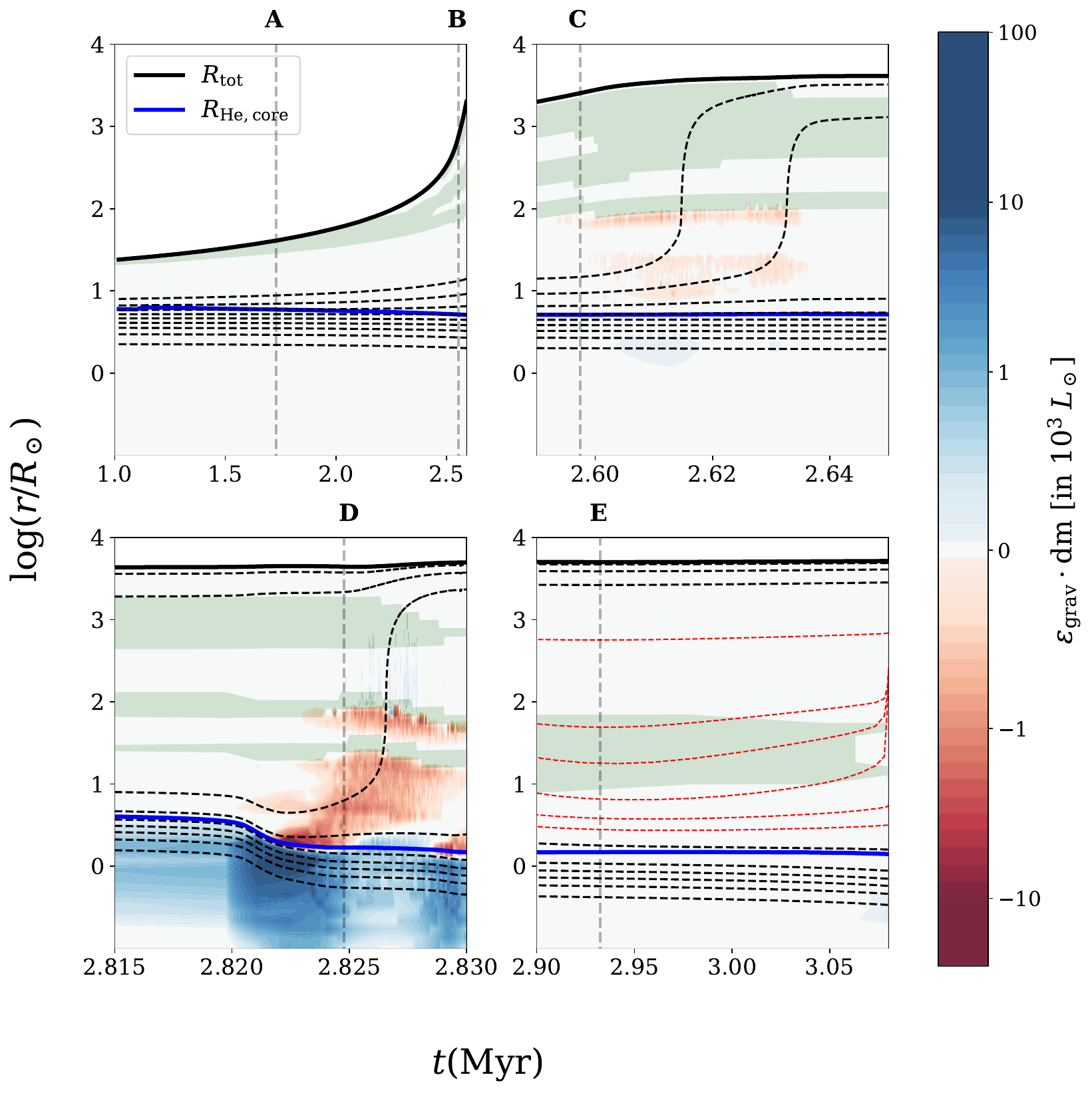}
    \caption{Radial Kippenhahn plot showing the internal profile evolution of the $100\,M_\odot$ model. The solid black and blue lines are the total and core radii respectively. {The black dashed lines track radius of constant $q = m/M_\mathrm{tot} = 0.1 - 0.9$ in increments of $0.1$. Additional constant $q$ lines between $q=0.6$ and 0.7 are also plotted in the bottom right plot.} Blue and red contours are used to show positive (contracting) and negative (expanding) contributions of the local gravitational source term, $\epsilon_\mathrm{grav}$. The green shading marks convective regions where $|1 - \Gamma_\mathrm{rad}| < 0.1$, and $\beta = P_\mathrm{gas}/P_\mathrm{tot} < 0.1$.  The four subplots correspond to (a) Inflation during the Main sequence phase (b) Expansion during the MS as the model loses thermal balance just below temperature of log($T_\mathrm{eff}$/K) $\sim3.8$ (c) Expansion once the core runs out of hydrogen (d) Inflating layers inside the model during core-He burning.  } 
    \label{fig: RT-time100_infl}
\end{figure*}

\subsection{Interplay between envelope expansion and inflation}
\label{sec: 100msun_exp_infl}

Based on the properties of inflation and expansion discussed till now, we can differentiate between the two phases on the evolutionary tracks of the 60 and $100\,M_\odot$ models. The $60\,M_\odot$ model inflates until log($T_\mathrm{eff}$/K) $\sim4$ while maintaining hydrostatic and thermal balance and does so stably. Just prior to core-H exhaustion, the model is inflated. As the core rapidly runs out of H, the model undergoes a total contraction phase, followed by H-shell ignition. This is followed by an expansion phase marked by thermal imbalance and instability (negative $W_{m}$) in the envelope.  The switch from inflation to expansion is clear in this case. 

The $100\,M_\odot$ model however, shows an interesting switch between inflation and expansion. On the surface, the evolution of the $100\,M_\odot$ model appears relatively straightforward -- the model evolves towards cooler temperatures till it becomes a RSG and then increases in luminosity. However, the internal radial variations of this model reveals the distinct inflation and expansion phases occurring during its evolution. To delve deeper into this model, we simultaneously look at the model's internal profiles and radial Kippenhahn diagrams in Figs. \ref{fig: infl_exp_profile_subplots} and \ref{fig: RT-time100_infl} respectively. 

Profiles are taken at five different points during the evolution of the $100\,M_\odot$ model marked from `A' to `E'. These correspond to the five columns in $\mathrm{Fig.}\,\ref{fig: infl_exp_profile_subplots}$. To complement these internal profiles, in $\mathrm{Fig.}\,\ref{fig: RT-time100_infl}$ {we present the internal radial evolution of the model as it evolves through the five points. This figure tracks the internal evolution of core radii, along with radii corresponding to constant $q = m/M_\mathrm{tot}$ values.} As we keep the total mass of the model unchanged, these lines can be thought of as tracing the movement of mass within the model. 

Layers that are out of thermal balance are marked with blue and red contours. Convective layers that have $|1-\Gamma_\mathrm{rad}| \,< \,0.1$ (close to unity) and the gas-to-total pressure ratio $\beta \,<\,0.1$ are shaded in green. We use a simple condition here which checks layers where high Eddington parameters are realized in inefficient convective regions where the flux transport is still dominated by radiation. Alternative criteria have been used previously with similar arguments as above, see for eg. \citet{Graf2011, Sanyal2015}. For a more detailed inflation criteria where the typical inflated morphology arises, see for eg. \citet{Grassitelli2018}.

\vspace{0.3cm}
\noindent
\textbf{1. Main sequence inflation}: The $100\,M_\odot$ model undergoes envelope inflation during its MS until log($T_\mathrm{eff}$/K) $\sim3.8$. We show internal profiles of points `A' and `B' during this inflation phase in the first two columns of $\mathrm{Fig.}\,\ref{fig: infl_exp_profile_subplots}$. Profile `A' is taken halfway through the MS, and profile `B' at log($T_\mathrm{eff}$/K) of 3.9. Both profiles have $\Gamma_\mathrm{rad}$ values very close to 1 in their inefficient convective zones.  This is the hallmark of inflated layers which have been allowed to settle in hydrostatic and thermal balance. Until log$(T_\mathrm{eff}/\mathrm{K}) \sim4.6$, inflation occurs in the convective zone about the Fe-bump at log($T$/K) $\sim5.3$. Below this temperature, the He\textsc{ii} opacity bump also becomes relevant. The $\Gamma_\mathrm{rad}$ remains close to unity, and the density flattening now occurs across both bumps. The quantity $W_{m}$ remains positive until log($T_\mathrm{eff}$/K) $\sim3.8$, albeit gradually reducing and approaching zero. 

The top left plot in $\mathrm{Fig.}\,\ref{fig: RT-time100_infl}$ shows the MS inflation phase. The model is in thermal balance as seen by the absence of red or blue contours. The outermost layers of the model have a green shading indicating their close proximity to the local radiative Eddington limit.  The total radius in this phase increases from 20 to $2000\,R_\odot$. In a relative sense, this is an increase by a factor of $\sim100$, showing the extent to which inflation affects the total radius. The green shaded region branches out at about 2.5 Myr due to the appearance of a new opacity bump -- the He\textsc{ii}-bump -- which further complicates the simple trunk-like density morphology picture as seen in the previous figure. The inflation is limited to the near-surface layers during the MS.

\vspace{0.3cm}
\noindent
\textbf{2. Thermal instability on the MS}: The model at profile `B' is clearly an inflated model with $\Gamma_\mathrm{rad}$ values close to unity across most of the envelope.  In this regard, we expect the layers to further inflate when the nuclear luminosity increases, as has been the case for most of the MS phase. However, inflation beyond this point results in an increase in the opacity and the sign of $W_{m}$ changes. The model undergoes a thermal instability runaway from log($T_\mathrm{eff}$/K) $\sim$ 3.8 to 3.5, as the $\epsilon_\mathrm{grav}$ in the envelope becomes increasingly negative.  This is an example of what occurs when an inflating envelope encounters a thermally unstable situation.  This is similar to the previously discussed runaway in the 15 and $60\,M_\odot$ models, but with a key distinction. In the 15 and $60\,M_\odot$ models, the runaway phase occurs after H is fully exhausted in the core. In the $100\,M_\odot$ case, the runaway phase already occurs on the MS as inflation brings the model to temperatures where the H-bump becomes prominent. 

Profile `C' is chosen halfway through this runaway phase. As seen in $\mathrm{Fig.}\,\ref{fig: infl_exp_profile_subplots}$ (column 3), the model has negative $W_{m}$ throughout the envelope, indicating that the actual luminosity can no longer match the changes in the nuclear luminosity. As seen in the HR diagram ($\mathrm{Fig.\,}\ref{fig: HRD_exp_condition_60_100}$), thick black borders eventually appear due to this growing difference between the luminosities,{ which is shown in a magnified inset plot.} Given that the model was in stable thermal balance just before entering this phase, the actual luminosity only diverges by a few percent from the nuclear luminosity before the model regains thermal balance as an RSG. Consequently, the red and blue $\epsilon_\mathrm{grav}$ contours seen in top right plot of $\mathrm{Fig.}\,\ref{fig: RT-time100_infl}$ during this phase are lightly shaded. The total radius in this phase increases from 2000 to $4000\,R_\odot$, doubling in a span of 0.05 Myr. The effect of this expansion can also be tracked internally, where the radius of constant q = 0.8 and 0.9 lines increase by two orders of magnitude. So from the definitions established in this work, the model \textit{does not inflate}, but instead \textit{expands} across the H-bump.

\citet{Sanyal2015} report maximum local $\Gamma_\mathrm{rad}$ values inside their models and find values as high as $\sim7$ when the models encounter the H-bump.  Here, we briefly discuss the phase the model is in when such high values are realized internally. During the runaway phase, the $\Gamma_\mathrm{rad}$ values can momentarily shoot up in the H-bump region. The model is precipitously moving away from thermal balance for the model to adjust to such sudden spikes in $\Gamma_\mathrm{rad}$.  A similar behaviour is seen in the $15\,M_\odot$ model as it evolves through its thermal instability phase, with $\Gamma_\mathrm{rad}$ values going up to $\sim2$. However, as the H-bump sweeps inwards towards higher densities, convection becomes more efficient and $\Gamma_\mathrm{rad}$ quickly reduces. 

The runaway phase ends at log($T_\mathrm{eff}$/K) $\sim3.5$ and the model regains thermal balance on the Hayashi line. The model first underwent a phase of MS inflation where only the outer-most envelope layers participated. This was followed by an expansion that redistributed tens of solar masses into the layers that had previously inflated, as indicated by the black dashed lines of constant $q=0.8$ and $0.9$. This is an example of an envelope that has been both \textit{inflated and expanded}.

The amount of mass in the high $\Gamma_\mathrm{rad}$ regions is orders of magnitude higher than before the runaway phase. Such high masses up to $\sim10^2\,M_\odot$ have been reported before especially at cooler temperatures \citep[see for e.g.][]{Sanyal2017}. The model finishes the rest of its core-H burning as an RSG evident by the first set of blue squares immediately following the runaway phase in $\mathrm{Fig.}\,\ref{fig: HRD_exp_condition_60_100}$ (in between points `C' and `D'). 

\vspace{0.3cm}
\noindent
\textbf{3. Core-H exhaustion}: As H depletes, the model once again goes out of thermal balance, and further expands along the Hayashi line. Profile `D' is chosen immediately after complete H exhaustion.  The quantity $W_{m}$ is negative in some parts of the envelope (cf. fourth column in $\mathrm{Fig.}\,\ref{fig: infl_exp_profile_subplots}$). The $\Gamma_\mathrm{rad}$ is very close to unity in the envelope, but the model is clearly out of thermal balance as evident by the red and blue $\epsilon_\mathrm{grav}$ contours in the bottom left plot of $\mathrm{Fig.}\,\ref{fig: RT-time100_infl}$. The $\epsilon_\mathrm{grav}$ contours here are significantly more pronounced compared to the top right plot, as the model goes completely out of thermal balance following core-H exhaustion. During this phase, the core contracts and the envelope layers simultaneously expand. There is a huge increase in the radius internally, for example, the radius of constant $q = 0.7$ increases from 10 to $2000\,R_\odot$, representing a 200-fold increase. In comparison, the total radius only increases from 4300 to $5000\,R_\odot$. Thus, expansion can occur internally without significant (relative) changes to the total radius of the model. 

\vspace{0.3cm}
\noindent
\textbf{4. Core-He burning}:
Following core-H exhaustion and an internal expansion episode, the model begins and ends stable core-He burning as an RSG. This is seen by the second set of blue squares in $\mathrm{Fig.}\,\ref{fig: HRD_exp_condition_60_100}$ (point `E'). The quantity $W_m$ is positive throughout the envelope and $\Gamma_\mathrm{rad}$ close to unity. At the onset of core-He burning, the density stratification displays a morphology that is a blend of inflated and expanded model properties -- a dense core with a steep density drop, a low-density envelope containing tens of solar masses within thousands of solar radii, and a trunk-like structure.

The bottom right plot in $\mathrm{Fig.}\,\ref{fig: RT-time100_infl}$ depicts the core-He-burning phase, where no blue or red contours are visible indicating thermal balance. Additional constant $q$ lines (in red dashed) are also included between $q = 0.6$ and $0.7$ to highlight radial evolution near the high $\Gamma_\mathrm{rad}$-low $\beta$ (green shaded) region. The constant $q$ lines near the high $\Gamma_\mathrm{rad}$ region gradually move outward. This is similar to inflation near the surface during the MS but occurs internally and is significantly weaker as the increase in luminosity during core-He burning is very small. The total radius of the model barely changes during this phase. Inflation can therefore occur both near the surface and deep within the model.

\vspace{0.5 cm}
To address the question of whether RSGs are expanded or inflated --  in order to achieve the radii and temperatures characteristic of RSGs, the model must cross the H-bump. The crossing of the H-bump can occur either after the MS, at which point the model is fully out of thermal balance and thermal instability drives it further from balance, or during the MS, where thermal instability initiates and causes the model to lose thermal balance. Regardless, all RSG envelopes have undergone a phase of thermal imbalance, and are thus expanded. Furthermore, for models with sufficiently high initial mass, inflation can occur during the MS, leading to RSGs that may exhibit a blend of both inflated and expanded morphology at the onset of core-He burning. 

\citet{Sanyal2015} found that models with an $M_\mathrm{init}$ of approximately $40\,M_\odot$ undergo inflation during the MS. Thus, for initial masses above roughly $40\,M_\odot$, RSG models at the onset of core-He burning would exhibit a blend of inflated and expanded morphology. However, this threshold also depends on processes that influence $\Gamma_\mathrm{rad}$ during the MS. For instance, if core overshooting mixing is more efficient during the MS, RSG models may develop the blended morphology at a lower initial mass.  The threshold mass for the blended morphology can be further lowered if $\Gamma_\mathrm{rad}$ increases during core-He burning, either due to an increase in $L_\mathrm{act}$ or a decrease in mass due to mass loss, allowing the models to undergo additional internal inflation.
 
In this section we employ simple evolution models devoid of any mass loss or rotation to distinguish between inflation and expansion. Equipped with this understanding, our aim in the following section is to explore important examples of inflation within models that change mass, either through winds or binary mass transfer, and discuss potential consequences for real stars.

\section{Discussion}
\label{sec: discussion}

Having presented the general properties of inflated and expanded models, we now discuss them in the context of specific stages of evolution of massive stars, namely Very Massive Stars (VMSs), red supergiants and their upper luminosity threshold, Heavy BHs and partially stripped stars located redward of the H-ZAMS.

\subsection{Vertical and Horizontal evolution of very massive stars}

In the previous section, our zero-mass-loss models with initial masses above $\sim60\,M_\odot$ experienced inflation during the MS. How would mass loss alter this picture? The decrease in mass would directly increase the radiative Eddington parameter (cf. Eq. \ref{eq: Eddington_parameter}), causing the models to inflate further. However, this holds true only up to a certain mass-loss rate, beyond which the reduction in mass begins to outweigh the effect of mean molecular weight on luminosity {(cf. $\mu$--effect from $\mathrm{Sect.}\, \ref{sec: redward_exp}$)}.

\begin{figure}
    \includegraphics[width = \columnwidth]{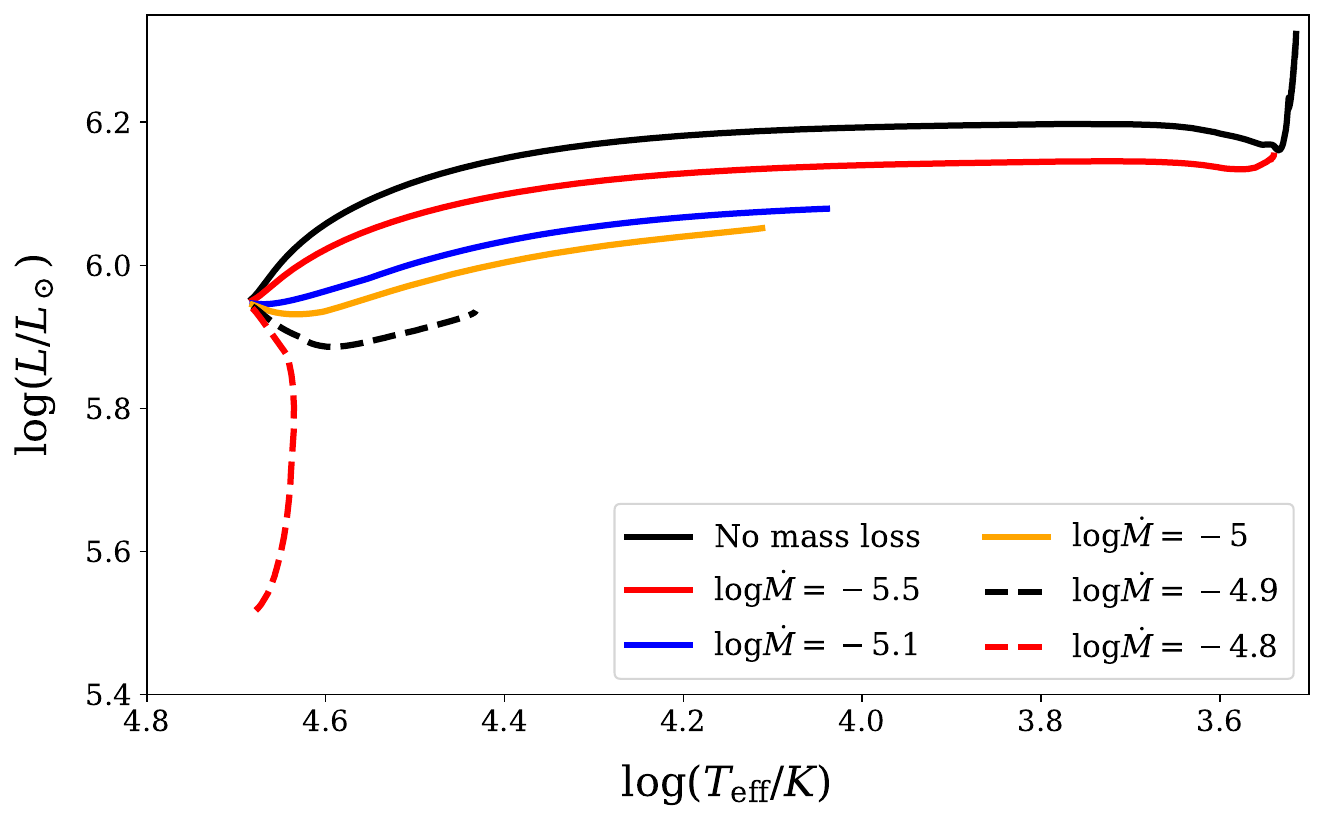}
    \caption{Hertzsprung-Russell  diagram tracks of our $80\,M_\odot$ model subjected to different mass-loss rates which are kept fixed during the MS. {All models finish core-H burning except the blue track which is terminated at $X_\mathrm{c}\sim 0.05$. Above a threshold mass-loss rate, the models no longer inflate, sharply changing behaviour and evolving vertically downward. }} 
    \label{fig: HRD_change_mass}
\end{figure}

If mass loss is strong enough to severely affect the size of the core and reduce the total luminosity itself, the increase in the Eddington parameter\footnote{We are talking about the surface Eddington parameter here. Near the surface, the radiative Eddington parameter is equal to the total Eddington parameter.} would be minimal, suppressing the inflation effect. Consequently, the model would evolve vertically downwards in the HR diagram. In $\mathrm{Fig.}\,\ref{fig: HRD_change_mass}$, we show this `flip' in the evolutionary behaviour from horizontal to vertical in our $80\,M_\odot$ model subjected to different constant mass-loss rates (see legend). For the $80\,M_\odot$ example shown here, the mass-loss rate where the flip occurs is about $10^{-4.9}\,M_\odot/\mathrm{yr}$. Above this mass-loss rate, the effect of increasing mean molecular weight is effectively negated and luminosity no longer increases during the MS, preventing inflation. The most massive stars therefore might not undergo inflation.

For low Eddington parameters, both the analytical CAK theory and semi-analytical Monte-Carlo (MC) models predict a shallow scaling of the mass-loss rate with the Eddington parameter \citep{CAK1975, Vink2000, Puls2008}. However, the MC models find a \textit{transition point} where the wind efficiency parameter $\eta = \dot{M}\varv_\infty/L_\mathrm{act,surf}/c$ crosses approximately unity, above which the mass loss scales steeply (with a power-law slope of $\sim5$) with the Eddington parameter \citep{Vink2011}. A similarly steep mass-loss scaling near the Eddington limit has been reported by \citet{GH2008} using the PoWR atmosphere code. A smooth change in this $\dot{M}-\Gamma$ slope has also been proposed through an extension to the CAK formalism by \citet{Best2020mdot}. The bottom line is that wind models indicate the mass-loss rate rapidly increases as the Eddington limit is approached. 

\begin{figure}
    \includegraphics[width = \columnwidth]{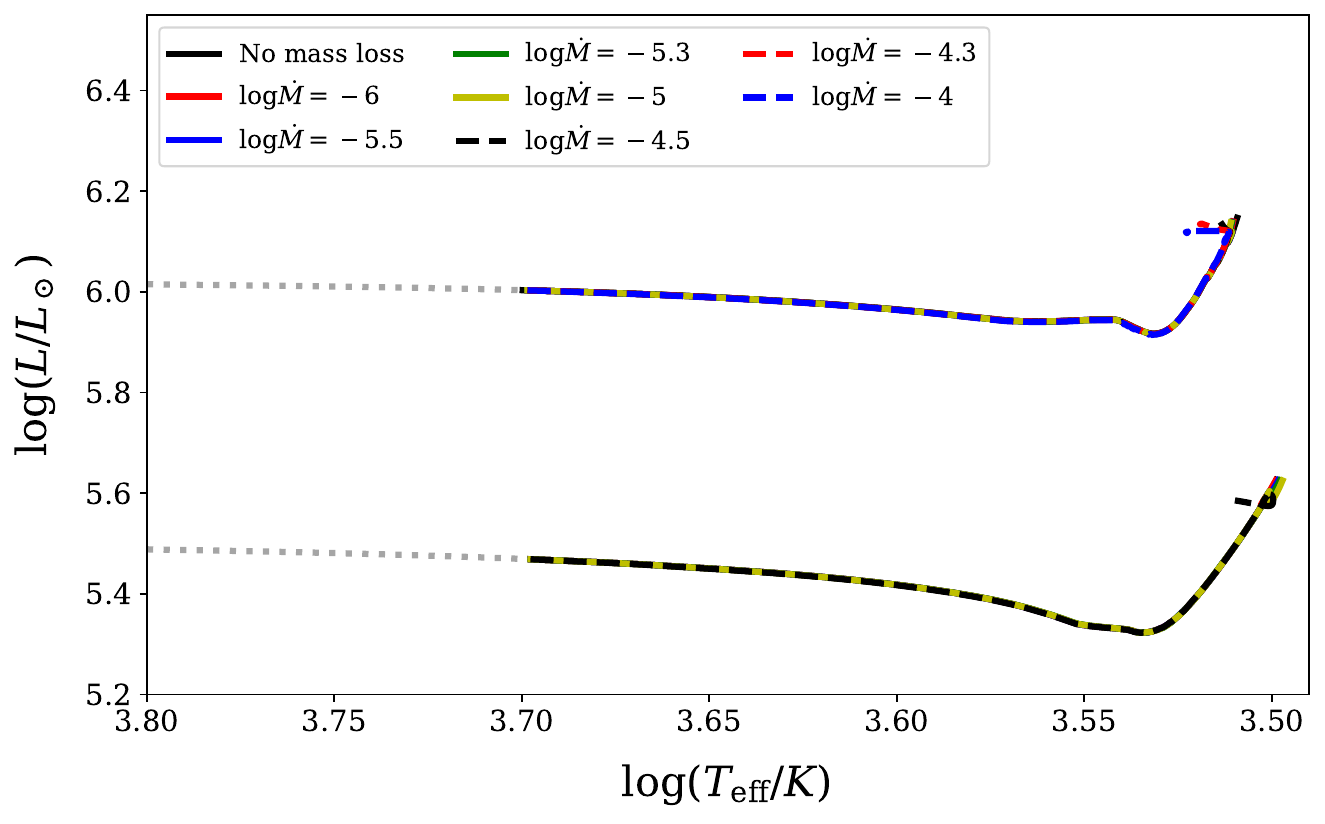}
    \caption{Hertzsprung-Russell  diagram tracks of our $30$ and $60\,M_\odot$ {model subjected to different mass-loss rates which are kept fixed} during the core-He-burning RSG phase. No mass loss is applied till log$(T_\mathrm{eff}) = 3.7$ below which the models are subjected to a constant mass-loss rate specified in the legend. {The luminosity of RSGs during core-He burning is primarily determined by the He core mass, which remains largely insensitive to the  mass-loss rate.
}} 
    \label{fig: HRD_change_mass_RSG}
\end{figure}

Recently, \citet{Sabhahit2022} implemented a steep scaling of mass loss above the transition point in VMS models up to $500\,M_\odot$, demonstrating a drop in luminosity. Such drop in luminosity and suppression of inflation have been previously documented in the literature using different codes and mass-loss inputs in the context of VMS models at high $Z$ \citep{Yung2008, Yusof2013, Kohler2015, Chen2015}. 

The population of WNh stars are our best testbeds to find empirical evidence to support the claims of wind models. WNh stars are hydrogen-rich WR stars of the `N' sequence with luminosities in excess of log$(L/L_\odot) \sim6$ \citep{Hamann2006, Martins2008, Crowther2010, Best2014}. They show all the signatures of an emission line-dominated spectra of a WR star but with non-negligible H, likely signifying that these objects are still core-H-burning objects. \citet{Graf2011} and \citet{Best2014} have investigated the mass-loss scaling in the Arches Cluster of our Galaxy and 30 Dor in the LMC respectively -- two young massive clusters containing O and WNh stellar populations. These studies empirically confirm that the mass loss of WNh stars scales steeply with the Eddington parameter.

In the Arches cluster, \citet{Vink2012} applied their model-independent method, known as the transition mass-loss point, to the O and WNh star population. Using this method, they estimate a mass-loss rate of $10^{-5.1}\,M_\odot/\mathrm{yr}$ based on an average luminosity of log$(L/L_\odot) \sim5.95$ for the transition O/WNh objects, which approximately corresponds to $80\,M_\odot$ on the H-ZAMS. This mass-loss estimate is roughly 0.2 dex lower than the threshold we find the `flip' in behaviour for our $80\,M_\odot$ model. If the mass-loss scaling with the Eddington parameter does indeed increase steeply above the transition point, then the most massive stars observed in the Galaxy and the LMC are likely evolving vertically downwards, which is qualitatively very different from the evolution of canonical massive stars (of $\lesssim 60\,M_\odot$). VMSs evolving vertically downwards also offer a simple explanation for the narrow range of observed temperatures of WNh stars across varying luminosities in both the Galaxy and the LMC \citep[see arguments in][]{Sabhahit2022}.

Beyond the Local Universe, individual objects cannot be resolved, so we lack empirical constraints on the WNh population. However, we can make stellar evolutionary predictions for the early Universe, when the metal content was likely lower than it is today. At low $Z$, the radiative force on metal lines is lower, leading to reduced mass-loss rates \citep{Abbott1982, Vink2001, Kudritzki2002}. This has been empirically verified, as OB-stars in the Galaxy exhibit higher mass-loss rates compared to their counterparts in the SMC, with a metal content of approximately one-fifth solar, for the same luminosity \citep{Massey2005, Mokiem2007}. Therefore, at sufficiently low $Z$, the mass-loss rates are not strong enough to prevent inflation if it occurs\footnote{For a rough estimate of the initial mass above which models begin to drop in luminosity at each metallicity, we refer the reader to Table 2 in \citet{Sabhahit2023}}. We emphasize `if it occurs' because the Fe-bump is significantly weaker at low $Z$, {causing the initial mass at which inflation is realized in models to shift upward \citep{Sanyal2017}}. However, strong inflation might still occur across the He\textsc{ii}-bump. Therefore VMSs in the early, metal-poor Universe could potentially inflate (and expand) until they become RSGs (cf. $\mathrm{Sect.\,}\ref{sec: Heavy_BH}$), where mass loss could occur due to eruptive events or turbulence \citep{Josselin2007, Kee2021, Cheng2024}.

\subsection{Upper luminosity limit for core-He-burning red supergiants}
\begin{figure*}
    \includegraphics[width = \textwidth]{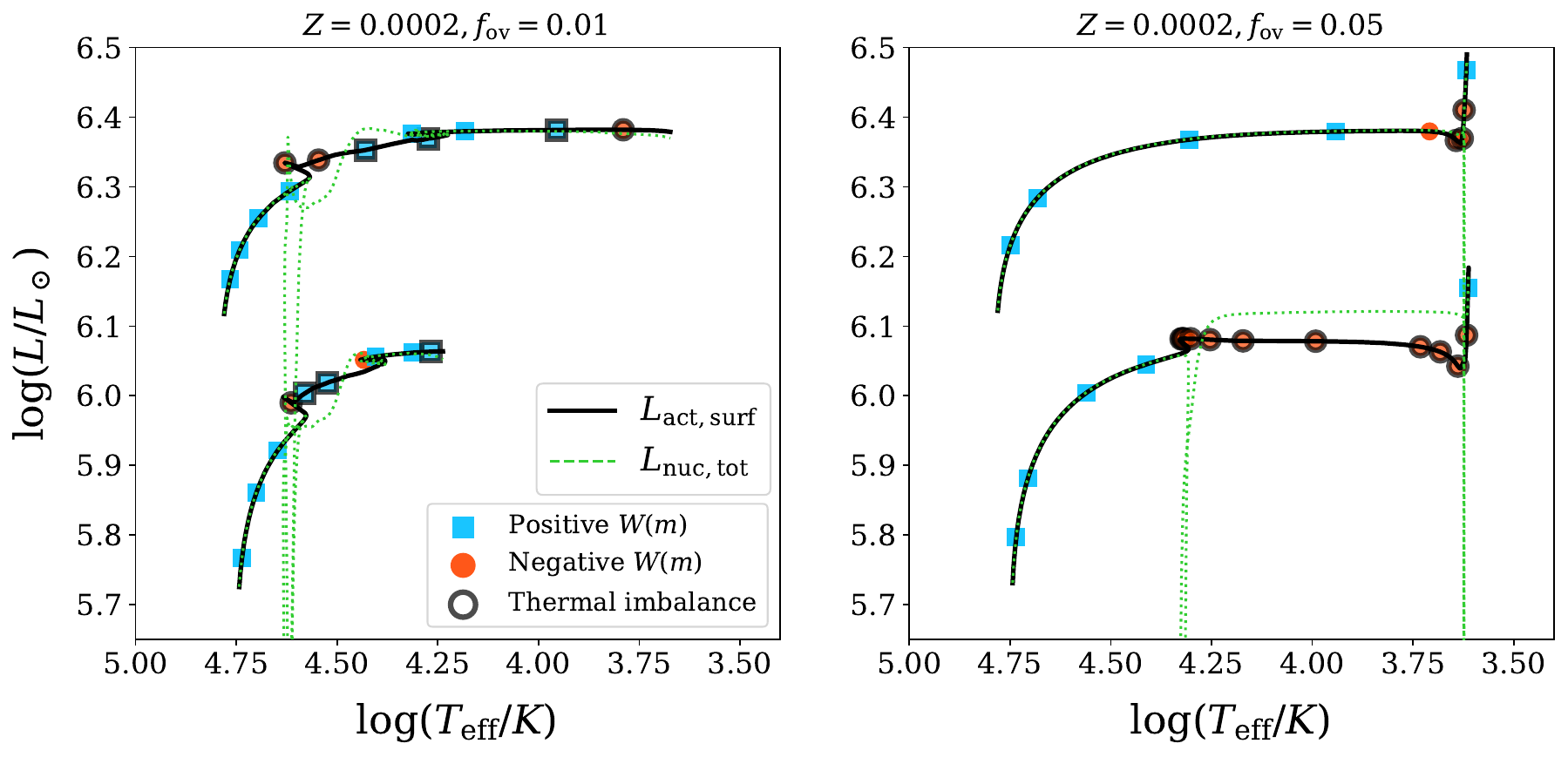}
    \caption{Hertzsprung-Russell  diagram tracks of 60 and $100\,M_\odot$ models at low $Z$ of $0.0002$ until core-He exhaustion. {The lines and coloured symbols have the same meaning as in $\mathrm{Figs.\,}\ref{fig: HRD_15os3}\;\mathrm{and}\;\ref{fig: HRD_exp_condition}$.} The left and right plot shows models with low ($f_\mathrm{ov} = 0.01$) and high ($f_\mathrm{ov} = 0.05$) core overshooting respectively. {With smaller core overshooting and low mass loss, models stabilize as compact BSGs during core-He burning.
}}

    \label{fig: HRD_low_Z}
\end{figure*}

Now we focus on our RSG models, particularly those that are burning He in their cores and H in a shell surrounding the He core. The 30 and $60\,M_\odot$ models are evolved with no mass loss until log$(T_\mathrm{eff}/K) = 3.7$, at which point both models have already formed an inert He core and are supported by a H-burning shell around the core. Below this temperature, we subject our models to different mass-loss rates that are fixed during the core-He-burning RSG phase and follow their HR diagram tracks during core-He burning, as shown in $\mathrm{Fig.}\,\ref{fig: HRD_change_mass_RSG}$. 

There is a stark difference between how models react to mass loss in the H-burning and the He-burning {RSG} phase. In the H-burning case, the model luminosities and temperatures immediately react to the input mass loss. The core size adjusts itself to the total mass. An H-burning model responds to strong mass-loss episodes with a sharp drop in luminosity, which can limit the increase in $\Gamma_\mathrm{rad}$ or even decrease it. Following strong or even eruptive mass loss, as recently suggested by \citet{Cheng2024} in the hot VMS regime, the models can stabilize at a lower luminosity, that is, the model can regulate its $L/M$-ratio if it approaches the Eddington limit by having a strong mass-loss episode and dropping its luminosity. 

This is however not the case for core-He-burning models. Once the He-core mass is set at the end of H-burning, it hardly reacts to the decrease in the mass of the H-envelope \citep[see also \textsc{snapshot} model tests from ][]{Farrell2020}. The luminosity and temperature remains constant while the mass of the star decreases. With a lower mass, it is easier for material to be ejected from the surface as the material has to be lifted out from a shallower potential well, likely increasing the mass loss. A runaway mass-loss process ensues where mass loss exponentially increases until a large fraction of the envelope is stripped. 

The actual amount of envelope lost will ultimately depend on the absolute mass-loss rates and how quickly they ramp up. If they ramp up quickly, the $\Gamma_\mathrm{rad}$ can reach unity throughout the entire envelope, as the model fails to regulate its $L/M$-ratio. For the test models subjected to the lowest mass-loss rates (below $\sim10^{-5}\,M_\odot/\mathrm{yr}$), the increase in $\Gamma_\mathrm{rad}$ during the core-He burning is negligible. These models begin and finish core-He burning as an RSG without any internally inflating layers. For the models subjected to the highest mass-loss rates, the  $\Gamma_\mathrm{rad}$ quickly approaches unity in the entire envelope, and a blended inflated+expanded morphology can be realised. 

If a large fraction of the RSG envelope is stripped, two things occur. First, the RSG models rapidly approach their radiative Eddington limit throughout the entire envelope and begin to internally inflate. A second effect of higher mass-loss rates is that a large fraction of the envelope is lost, and the model begins to evolve bluewards. As the model attempts to cross the H-bump the quantity $W_m$ becomes negative in the envelope. This scenario is similar to the first crossing of the H-bump from blue to red in the $100\,M_\odot$ model, where a stable, inflated envelope suddenly encounters the strong H-bump {(cf. evolution through points `B' to `C' in $\mathrm{Fig.}\,\ref{fig: HRD_exp_condition_60_100}$)}. In the present case however, the models attempt to cross the H-bump from red to blue. Numerical problems arise with models requiring smaller and smaller timesteps to resolve the evolution and the model evolution is terminated. While we are unable to follow the evolution of these models until core-He exhaustion, it is likely that they strip their envelopes, expose their hot He cores, and end up near the He-ZAMS.

The location in the HR diagram where such exponentially increasing mass loss and high $\Gamma_\mathrm{rad}$ values in the entire envelope is predicted to occur is observed to be devoid of any objects in Nature. \citet{HD1979} noted an observed absence of cool supergiants above a certain bolometric magnitude both in the Galaxy and the LMC. The upper luminosity limit for the population of RSGs is well established across multiple galaxies and corresponds to log$(L/L_\odot) \sim 5.5$ or initial masses of about $25-30\,M_\odot$ \citep{Massey2016, Davies2018, Davies2020, McDonald2022}. The general consensus for the existence of such a limit in the first place is attributed to strong mass loss which could be due to violent radiative or mode-resonance instabilities \citep{Lamers1988, Glatzel1993, Sanyal2015, Vink_Sabhahit2023, Fuller2024, Cheng2024}. The combination of strong radiative instabilities in the envelope and the failure of our models to regulate their $L/M$-ratios could result in strong mass-loss events, potentially explaining the sharp cutoff in the observed luminosity distribution of the coolest supergiants. However, alternative explanations involving mixing processes have also been proposed to explain the luminosity limit \citep[see for e.g. ][]{Abel2019, Higgins2020, Gilkis2021, Sabhahit2021}.



\subsection{Heavy black holes at low $Z$}
\label{sec: Heavy_BH}

In this section we investigate the effects of having lower metal content on the expansion and inflation phases in our models. We run 60 and $100\,M_\odot$ models at a hundredth solar metallicity ($Z = 0.0002$) with both low and high core mixing efficiency ($f_\mathrm{ov} = 0.01, 0.05$). Their HR diagram tracks {until core-He depletion} are shown in $\mathrm{Fig.}\,\ref{fig: HRD_low_Z}$. Metal content in the envelope can affect our models in two important ways. First, at lower $Z$ the models are hotter due to an opacity effect. One can directly compare the high overshooting models here (right plot) with the models in $\mathrm{Fig.}\,\ref{fig: HRD_exp_condition_60_100}$. This includes all important phases: H-ZAMS, TAMS location and Hayashi line. Second, the $\Gamma_\mathrm{rad}$ around the Fe-bump changes with $Z$. In the high $Z$ case, the $\Gamma_\mathrm{rad}$ is very close to unity, and the 60 and $100\,M_\odot$ models show a clear inflated morphology already on the H-ZAMS. At low $Z$, the 60 and $100\,M_\odot$ models on the H-ZAMS are not inflated and do not exhibit any trunk-like morphology in their envelope. The initial mass where models show inflated morphology on the H-ZAMS is shifted to higher values for lower $Z$ \citep[see models of ][]{Sanyal2017}.

Evolving off the H-ZAMS, the models with higher core overshooting show signs of inflation albeit towards the end of their MS across the He\textsc{ii}-bump. {The $100\,M_\odot$ model inflates across the He\textsc{ii}-bump, and then expands across the H-bump, becoming core-H burning RSGs. Such very high luminosity RSGs have been investigated previously in the context of very low $Z$ \citep{Yoon2012, Szecsi2015}, although such extended envelopes could potentially be destroyed in pulsations or eruptive events \citep{Moriya2015, Cheng2024}.}

For lower convective overshooting, the core size is smaller which limits the increase in luminosity during the MS and envelope inflation is suppressed. Immediately following the core-H exhaustion, the model goes out of thermal balance as seen by the appearance of red circles with thick black border. However, the model quickly regains thermal stability since the temperature is too hot for the envelope opacity to increase upon local expansion, i.e. the model is too hot for the opacity effect described in $\mathrm{Sect.\,}\ref{sec: redward_exp}$ to take place. The models stabilize as a BSG and finish their core-He burning there.  

Recently, multiple BHs from Gravitational wave events have their pre-merger masses estimated to be in the traditional pair-instability (PI) gap between $M_\mathrm{BH} \approx 50$ and $130 \,M_\odot$ \citep{Abbott2020, Abbott2024}. The PI gap is a gap in the mass distribution of heavy BHs due to the occurrence of pair-instabilities above a certain initial mass when the temperature in the core becomes hot enough to cause pair (electron-positron) production. This softens the radiation pressure dominated equation of state causing a collapse and subsequent catastrophic explosion completely disrupting the star leaving behind no remnant \citep[][for a recent review]{Barkat1967,  Rakavy1967, Glatzel1985, Fryer2001, Woos07, Wooseley2017, Marchant2019, Leung20, Farag2022, Renzo2024}. Several explanations have been proposed to account for this disparity in the BH mass distribution, including low $Z$, reduced $^{12}\mathrm{C}(\alpha, \gamma)^{16}\mathrm{O}$ reaction rates, smaller core sizes resulting from either low convective boundary mixing or post-collision remnants, among others \citep{Spera2019, Farrell2021_BH, Vink2021, Costa2022}.

The lack of inflation and negligible mass loss in the low $Z$, low overshooting scenario make them the most promising candidates for forming heavy stellar BHs in the traditional pair-instability mass gap. These models stay blue for the entirety of their core-H and core-He-burning phase, that is, they do not become luminous RSGs possibly avoiding strong mass loss or eruptive events \citep{Tanikawa2021, Vink2021}. The low $Z$ also ensures that the line-driven mass loss during the MS and the subsequent core-He-burning phase is negligible as there is less metals to drive a steady wind.

The low overshooting predominantly prevents expansion beyond the MS, while the low $Z$ limits inflation during the MS. Together, these factors work to stabilize the models as BSGs. {Compared to RSG envelopes, the smaller radius and compact nature of BSGs means that only a small fraction of the envelope is lost through ejecta in case of a failed supernova. Recent estimates by \citet{Fernandez2018} predict ejected masses of only $\sim0.2\,M_\odot$ from a BSG progenitor, while a large fraction of the envelope could potentially end up in the BH.}

{Additionally}, with low overshooting, the core size is small enough to prevent the occurrence of pair instabilities. Although this phenomenon can still occur, a smaller core for a given initial mass shifts the threshold mass at which it happens to higher values, potentially leading to the formation of the heaviest stellar BHs \citep{Winch2024}.

{A similar structural configuration of a small core mass relative to a large total mass can also be achieved through a post-MS merger process.} Once a He core has formed, it remains largely unaffected by changes to the star's total mass (as discussed in the previous section). The H-rich material from the secondary star is added to the envelope of the primary, with only a small fraction of the mass lost during the merger \citep[for details, see][]{Costa2022}. The end-product has a similar single-star configuration characterized by a low core-to-total mass ratio, with the models ultimately stabilizing as BSGs.

\subsection{`Puffed-up' envelopes of post-interacting binaries}
\label{sec: binary}
\begin{figure*}
    \includegraphics[width = \textwidth]{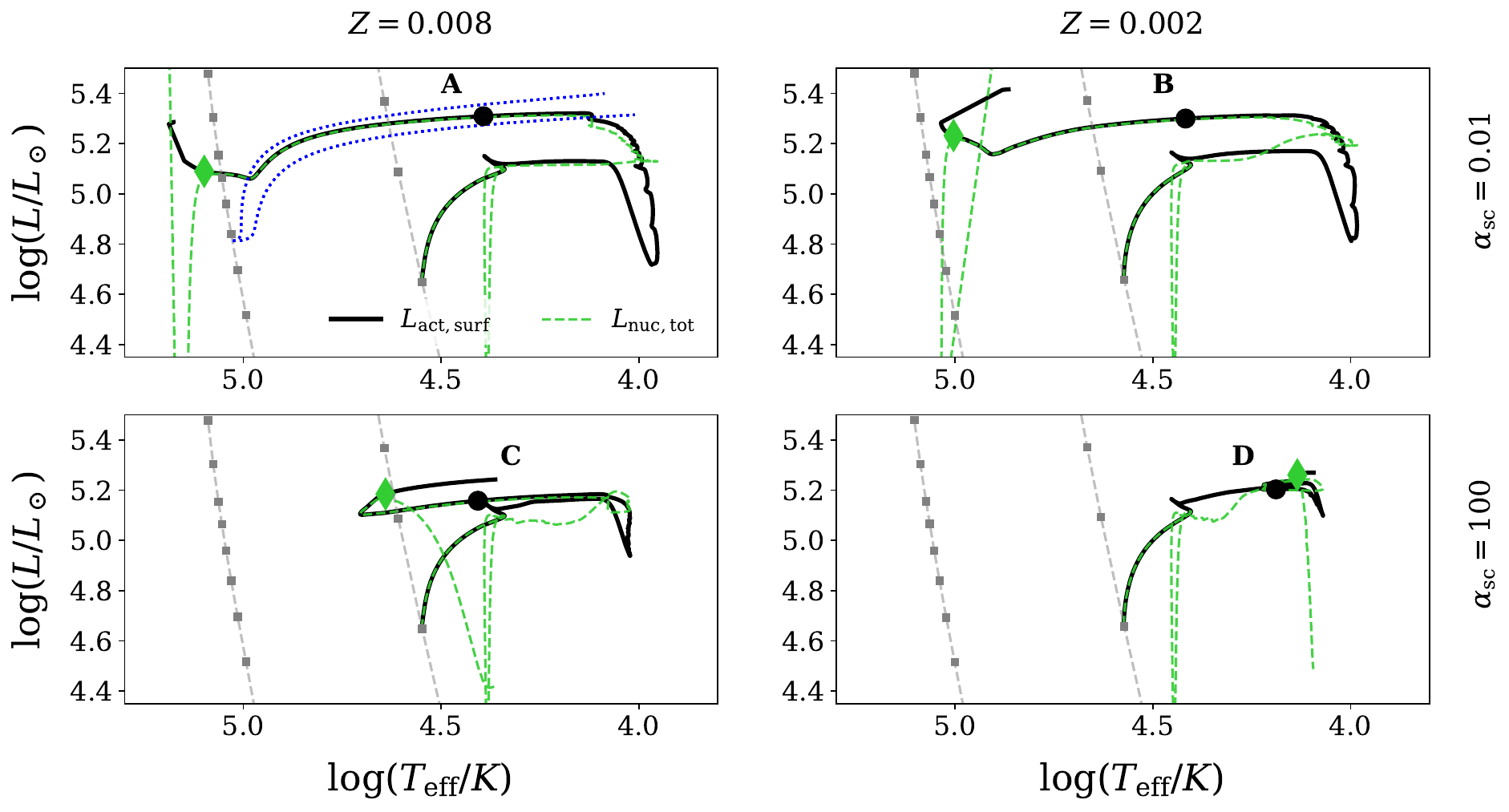}
    \caption{{\textit{Left:} Binary evolution tracks of a $20\,M_\odot$ primary star at $Z = 0.008$ and \textit{right:} at $Z = 0.002$ undergoing thermal mass transfer during the expansion phase. The two rows correspond to different value of semi-convection parameter used ($\alpha_\mathrm{sc} = 0.01$ and $100$). The colours of the binary tracks have the same meaning as in $\mathrm{Fig.\,}\ref{fig: HRD_15os3}$. The green diamonds mark end of core-He burning. The two grey dashed lines are the H- and He-ZAMS with different initial masses marked as grey squares. Black circles (A-D) are marked where the models regain thermal balance after the MT phase. The two blue dotted curves show loci of structure models with increasing H envelope mass on top of a $7\,M_\odot$ pure-He core, corresponding to the different mass fraction of H added: $X = 0.1, 0.2$. 
    These structure models are over-luminous compared to their He-ZAMS counterparts of the same mass, and illustrate the inflated nature of models with a low-mass, H-poor envelope above a He-burning core.
    }}
    \label{fig: puffed_env_binary}
\end{figure*}

\label{sec: binary}
\begin{figure*}
    \includegraphics[width = \textwidth]{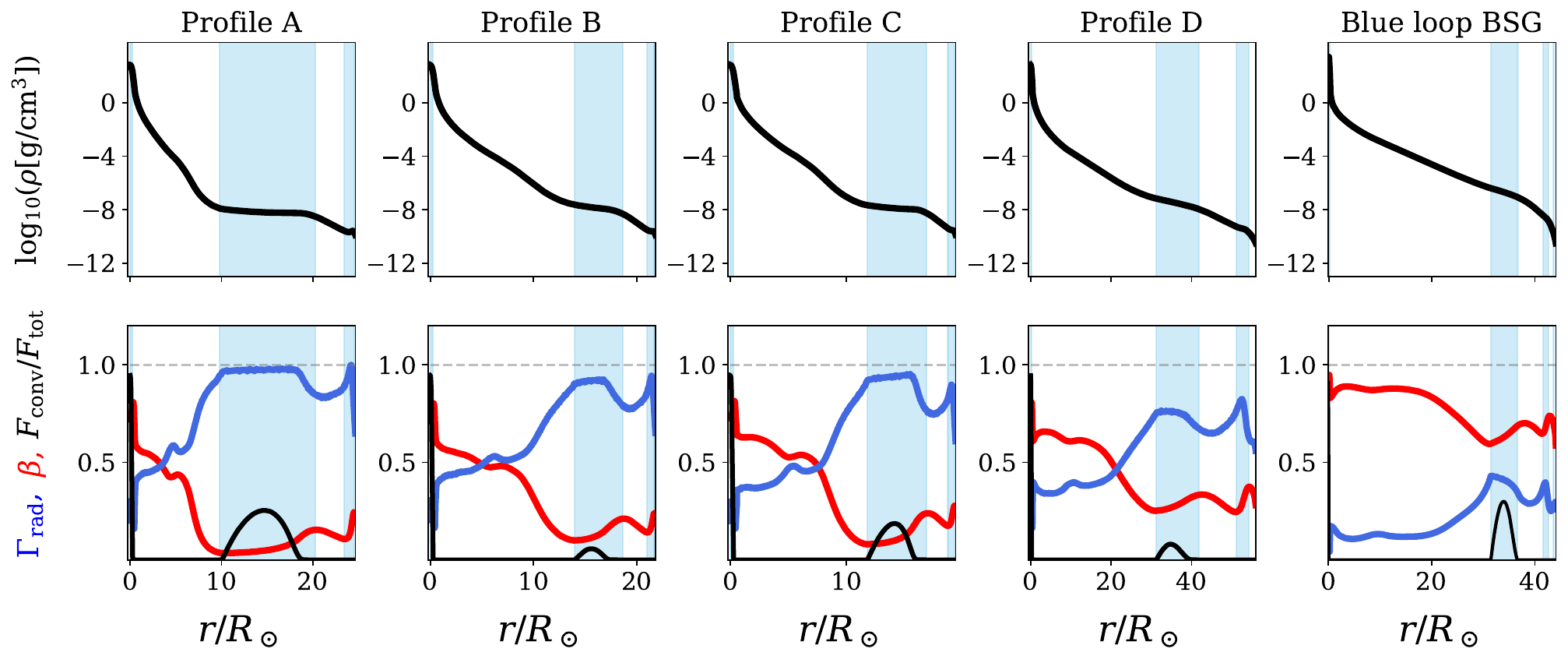}
    \caption{{Stratification of density (top row), local $\Gamma_\mathrm{rad}$, gas-to-total pressure ratio $\beta$ and convective flux fraction (bottom row). Five profiles are shown: four marked points (A-D) from $\mathrm{Fig.}\,\ref{fig: puffed_env_binary}$ and a blue loop BSG profile from $\mathrm{Fig.}\,\ref{fig: blue_loop}$. The blue shaded region indicates convectively unstable layers.}} 
    \label{fig: puffed_env_binary_profiles}
\end{figure*}

In this section we examine models consisting of a low-mass, H-poor envelope over a He-burning core. {This configuration is frequently referenced in this section, so we shorten it to `H-poor env + He-core' configuration.} These models exhibit higher luminosity compared to their pure-He counterparts of the same mass, potentially bringing them closer to their radiative Eddington limit. One evolutionary pathway to achieving this configuration involves envelope stripping in a single-star scenario as a result of substantial mass loss during the RSG phase. 

Alternatively, in binary systems, stable mass transfer (MT) via RLOF, occurring either during the MS or the post-MS expansion phase, may strip the envelope of the primary star \citep{Paczynski1967, Kipp1969, Podsiadlowski1992, Wellstein2001, Petrovic2005, Eldridge2008, Laplace2021, Klencki2022}. {A In particular, the post-MS expansion phase MT scenario can result in an `H-poor env + He-core' configuration.} 

Whether post-RLOF models fully strip or not depends on the envelope mass remaining after MT and the post-RLOF wind strengths  \citep{Gotberg2017, Gilkis2019, Laplace2020, Klencki2022}. {If the primary rapidly expands after core-H exhaustion, with the next stable configuration only achievable as an RSG, it quickly fills its Roche lobe and loses most of its envelope through thermal timescale MT. As a result, the post-RLOF model is left with a low-mass envelope relative to its core, containing very little hydrogen.} Subsequent wind-driven mass loss could lead to complete stripping of the primary. 

{In contrast, if the primary can stabilize and begin He burning as a compact BSG, it may undergo a brief phase of thermal timescale MT with partial envelope stripping, or MT on a thermal timescale followed by a slower nuclear timescale \citep{Klencki2022}. Should this envelope survive post-RLOF wind mass loss, the model remains partially stripped -- sometimes cooler than both the He- and H-ZAMS -- throughout the core-He burning phase. Factors like high semi-convection, low metallicity, or reduced overshooting, which stabilize the primary as a compact BSG, can result in such partially stripped configuration \citep[for more details, see series of papers by][]{Klencki2020, Klencki2022, Dutta_Klencki2024}.}

Recently, \citet{Ramachandran2023} identified the first partially stripped massive star in a binary system with a Be companion in the SMC. Three additional partially stripped + Be/Oe systems have been identified in the Clouds using data from the X-Shooting ULLYSES (XShootU) project \citep{Ramachandran2024}. Similarly, \citet{Villaseor2023} discovered a partially stripped star with a B-type companion in the LMC. Previously proposed BH + Be systems, most notably LB-1, have been reinterpreted as partially stripped + Be systems \citep{Bodensteiner2020, Shenar2020}. In the literature, such partially stripped systems have been referred to by various terms such as `bloated' or `puffed-up' stars \citep{Bodensteiner2020, Ramachandran2024, Dutta_Klencki2024}. 

{This section focuses on the internal profiles of such `puffed-up' stars, whether they represent a brief transitional phase or remain throughout core-He burning in a partially stripped configuration. In $\mathrm{Fig.}\,\ref{fig: puffed_env_binary}$, we present HR diagram tracks of binary models at $Z = 0.008$ and $0.002$, and two different values of the semi-convection parameter, $\alpha_\mathrm{sc} = 0.01$ and $100$.  The initial mass of the primary is $20\,M_\odot$ with a mass ratio of $q=0.6$. The initial period is set to $50$ days. } 

{Points A–D are marked where the models regain thermal balance after the MT phase, and their internal density profiles are shown in $\mathrm{Fig.}\,\ref{fig: puffed_env_binary_profiles}$. For comparison, profile of a BSG during the blue loop phase from $\mathrm{Fig.}\,\ref{fig: blue_loop}$ is also shown. One immediately noticeable trend is the higher $\Gamma_\mathrm{rad}$ in the post-RLOF profiles compared to the blue loop BSG profile. The stripping of the envelope causes the He-rich core to occupy a larger fraction of the star, leading to a higher $L_\mathrm{rad}$ and $\Gamma_\mathrm{rad}$. For example, at point `A' the model has a core-to-total mass ratio of $0.84$ and a surface luminosity of log$(L/L_\odot) = 5.3$. The blue loop BSG model has a core-to-total mass ratio of $0.24$ and a surface luminosity of log$(L/L_\odot) = 4.26$.}

{The chosen values of $Z$ and $\alpha_\mathrm{sc}$ parameters allow us to investigate the different internal profiles of `H-poor env + He-core' configurations with varying core-to-total mass ratios and envelope H mass fractions:} \\

\noindent
{1. For $\alpha_\mathrm{sc}=0.01$ at $Z=0.008$, we observe full envelope stripping. After the MS, the model rapidly expands quickly filling its Roche lobe. The primary's envelope is nearly fully stripped in a thermal timescale MT phase, leaving behind a low mass, H-poor envelope surrounding a He-rich core. For example, once the model regains thermal balance after the RLOF phase ends -- marked by a black dot as point `A' -- it has a small envelope of $\sim1.46\,M_\odot$ on top of a $\sim7.58\,M_\odot$ pure-He core with a low mass-averaged H mass fraction of $0.166$ in the envelope. The model has begun stable core-He burning redward of the H-ZAMS, a stage referred to as the `puffed-up' stripped star (PSS) phase by \citet{Dutta_Klencki2024}.}
    
{We show the internal density profile at point `A' during the PSS phase in $\mathrm{Fig.}\,\ref{fig: puffed_env_binary_profiles}$. The $\Gamma_\mathrm{rad}$ is close to unity across both the Fe and He\textsc{ii}-bumps. The gas-to-total pressure ratio $\beta$ is close to zero. The density gradient flattens across both bumps in layers with high $\Gamma_\mathrm{rad}$, meaning that this `H-poor env + He-core' configuration exhibits all the characteristics of an inflated model.}

\vspace{0.5 cm}
\noindent
{2. For $\alpha_\mathrm{sc}=0.01$ at $Z = 0.002$, the model begins core-He burning redward of the H-ZAMS, similar to the previous case. However, there are two small differences. First, the envelope is slightly more massive at $1.94\,M_\odot$ on top of a $\sim7.65\,M_\odot$ pure-He core with a higher H mass fraction, averaging $0.21$. Second, due to the lower metallicity, the post-RLOF model is not fully stripped and remains cooler than the He-ZAMS throughout the core-He burning phase.}

{A similar evolution is also realised in the $\alpha_\mathrm{sc}=100$ model at $Z=0.008$ -- the model begins core-He burning cooler than the H-ZAMS, but does not fully strip and remains cooler than the He-ZAMS.}

{Internal profiles at `B' and `C' during the PSS phase show high $\Gamma_\mathrm{rad}$ values across both the Fe and He\textsc{ii} bumps of roughly 0.9, while the $\beta$ parameter is close to 0.1. Although $\Gamma_\mathrm{rad}$ is not strictly unity, we see the flattening of the density gradient in these layers, once again displaying characteristics of inflated layers.}

\vspace{0.5 cm}
\noindent
{3. For $\alpha_\mathrm{sc}=100$ at $Z = 0.002$, the combination of strong semi-convective mixing and low $Z$ favour a BSG configuration after core-H exhaustion. The model is only partially stripped during the MT phase, leaving behind an envelope that is significantly more massive and richer in H compared to the previous cases. For example, at point `D' the model has a $4.4\,M_\odot$ envelope on top of a $7.6\,M_\odot$ He core with an averaged envelope H mass fraction of $0.46$.}

{The internal profile at point `D' reflects this lower core-to-total mass compared to the previous cases by having a lower $\Gamma_\mathrm{rad}$ (of $\sim 0.75$). The flattening in the density gradient is barely noticeable, resulting in a profile more akin to the BSG configuration during the blue loop phase of the $10\,M_\odot$ model discussed in $\mathrm{Sect.}\,\ref{sec: redward_exp}$. This model no longer has a `H-poor env + He-core' configuration and is not inflated.}

\vspace{0.5 cm}
To further illustrate the inflated nature of such {`H-poor env + He-core'} configurations, we construct structure models in reverse. That is, we slowly add H on top of a pure He-model of {$7\,M_\odot$, which roughy corresponds to the core mass of our binary models.} The accretion rate is $10^{-12}\,M_\odot/yr$, which is slow enough for the model to maintain hydrostatic and thermal balance, thus {generating a locus of structure models with a constant He core mass and varying envelope mass.} This procedure is similar to the \textsc{snapshot} models from \citet{Farrell2020}. Material is continuously accreted onto the pure He model without altering {its internal composition, achieved by turning off chemical compositional changes due to nuclear burning and mixing processes in MESA. Two such loci are shown corresponding to the different amounts of H mass fraction in the added material ($X=0.1$ and $0.2$). }

The added {material} begins burning H in a shell around the He-core, causing the luminosity to increase rapidly, with the H-burning shell becoming the dominant source of luminosity. The luminosity increases by approximately 0.5 dex during the experiment. Meanwhile, the mass accreted is on the order of a solar mass. This significant increase in the luminosity-to-mass ratio results in $\Gamma_\mathrm{rad}$ crossing unity in the Fe-bump.  Consequently, the structure models develop inflated morphology to maintain $\Gamma_\mathrm{rad}$ by reducing the density across the Fe-bump. The added H increases the opacity throughout the envelope, contributing to the radius expansion as $L_\mathrm{act}$ decreases, leading to a local expansion. However, the material added to the top of the He core in our tests is rich in He, so the increase in opacity is relatively small \citep[cf. tests in ][]{Farrell2020}.

{The evolution of the primary with an `H-poor env + He-core' configuration after MT closely traces these loci in reverse. As H burns in the shell, the He core gradually grows decreasing the envelope mass. This is further aided by wind mass loss which influences the duration of core-He burning spent in the transitory phase and determines whether the model reaches the He-ZAMS, as the envelope must be fully stripped for this to occur.} As the H shell becomes thinner, the total luminosity of the model begins to decline {because the total luminosity is dominated by the H-shell luminosity. }Although the contribution from the He core to the total luminosity increases, the declining luminosity from the H shell due to the thinning envelope results in an overall drop in total luminosity. Consequently the $\Gamma_\mathrm{rad}$ reduces. Since this phase exactly mirrors inflation but in reverse, we call it \textit{deflation}. Throughout the deflation phase, the model remains in thermal balance. This phase has been called `contraction' before \citep[for eg.][]{Dutta_Klencki2024}, but the removal of inflated layers as $\Gamma_\mathrm{rad}$ reduces should more appropriately be termed deflation.

Deflation of an {`H-poor env + He-core' configuration} can also occur at low $Z$ even when wind mass loss is weaker. {The deflation phase still occurs at low $Z$ because mass loss is not the only mechanism that thins out the H envelope. The gradual increase in the He core} due to H-shell burning leads to a corresponding decrease in the relative envelope mass, which lowers $\Gamma_\mathrm{rad}$ and causes the envelope to deflate. {In fact, \citet{Dutta_Klencki2024} find that H-shell burning has a more dominant effect compared to wind mass loss in the concerned mass range. }

{In summary, model configurations with a low-mass, H-poor envelope above a He core exhibit all the properties of an inflated model. As the H-poor envelope gradually thins -- either by H shell burning or wind mass loss -- the model \textit{deflates} and adjusts to the decreasing $\Gamma_\mathrm{rad}$. Such configurations can be realized in post-RLOF models that have undergone stable MT on a thermal timescale during the rapid envelope expansion phase following the primary's core-H exhaustion and can be understood using simple inflated single-star configurations.}


\section{Overview and conclusions}

The aim of the present paper is to elucidate the differences between envelope expansion observed in stellar evolution models beyond the MS and the phenomenon known as inflation, which occurs when high Eddington parameters are realised within hydrostatic stellar models, {and can therefore occur already on the MS.} While the increase in stellar radii by two to three orders of magnitude across the Hertzsprung gap has been well documented in the literature for decades, the concept of internal layers undergoing radial adjustment to regulate internal Eddington parameters is a more recent development. Although both phenomena can lead to a rapid increase in stellar radius, the underlying properties of the envelope in each case are fundamentally different. Furthermore, this paper revisits the terminology previously used to describe very low-density, tenuous envelope layers, highlighting {instances in stellar evolution} where these distinct phenomena can be easily conflated or misnamed. By clarifying this terminology, we aim to {reduce confusion and provide a clearer framework for future studies.}

We begin by clearly distinguishing envelope expansion ---the dwarf-to-giant transition typically seen in post-MS models--- from the phenomenon of inflation. The expansion phase is characterised by thermal imbalance, where the nuclear energy produced in the core does not match the energy lost from the surface. This discrepancy is compensated for by a gravitational source term, $\epsilon_\mathrm{grav}$, which is crucial for understanding how surface properties such as luminosity and temperature respond to internal energy and chemical changes within the model. The term $\epsilon_\mathrm{grav}$ represents the excess energy that cannot be easily dissipated, leading to local expansion, or the energy deficit that causes layers to contract in response. 

When a mismatch arises between the actual and nuclear luminosity, and if the local expansion or contraction tends to reduce this discrepancy, the model remains in (or moves towards) thermal balance. Conversely, if the adjustment causes the model to stray further away from thermal balance, then the model precipitously moves away from thermal balance. This can occur if local expansion (contraction) causes the opacity to increase (decrease) in the envelope. The actual luminosity can no longer match the increase (decrease) in the nuclear luminosity leading to an unstable scenario. 

The expansion phase, typically seen beyond the MS phase in stellar models, includes a period of thermal instability as the model crosses the H-bump. We also identify cases where the H-bump is encountered during the MS itself. In such scenarios, models lose thermal stability when crossing the H-bump and eventually move out of thermal balance. Thermal balance is eventually restored when a deep convective layer forms towards the end of the expansion phase. During blue loop episodes, thermal instability is triggered once again while crossing the H-bump. This transition from a red-to-blue supergiant configuration mirrors the earlier expansion phase in reverse. The entire phase of evolution where the model increases (or decreases) in radius on a thermal timescale and remains out of thermal balance with non-negligible negative (positive) $\epsilon_\mathrm{grav}$ is referred to as expansion (or contraction) in this work.

In contrast, inflation refers to the phenomenon where stellar layers adjust themselves to very low densities, effectively reducing the opacity to avoid a super-Eddington condition inside the model. A trunk-like morphology is realised in such inflated layers, with flattening in the density and gas pressure gradient. Hydrostatic and thermal balance is retained in inflated layers; that is, inflation can already occur on the MS. The outward radiative force in inflated layers is very close to the gravitational force, the ratio of which is called the radiative Eddington parameter $\Gamma_\mathrm{rad}$. The $\Gamma_\mathrm{rad}$ can locally exceed unity, and this is accompanied by a positive gas pressure gradient to balance the forces. The gas-to-total pressure is close to zero in inflated layers. 

Inflated layers can already form on the H-ZAMS for initial masses above $\sim60\,M_\odot$ at Galactic $Z$, and can cause the H-ZAMS to bend towards cooler temperatures. During the MS, the luminosity increases as H fuses into He, which increases the $\Gamma_\mathrm{rad}$ parameter.  Layers further inflate to accommodate the increase in $\Gamma_\mathrm{rad}$. On the MS, inflated layers can
even appear at lower initial masses due to this effect. \citet{Sanyal2015} find inflated layers already present in their $40\,M_\odot$ models during the MS. If such models end up as RSGs, they exhibit a blend of inflated and expanded morphology.

The inflation effect can also be influenced by processes that change $\Gamma_\mathrm{rad}$. For example, enhanced core-boundary mixing can indirectly trigger inflation during the MS;  even for stars with lower initial masses. Similarly, strong mass loss during the MS can cause the luminosity to stop increasing and instead drop sharply. Inflated layers are destroyed in such a scenario and the model evolves vertically downward in the HR diagram.

During core-He burning, models lose their luminosity sensitivity to the total mass. Once the He core mass is established at the end of core-H burning, it remains largely unaffected by changes in the mass near the surface. This has implications for supergiant configurations and envelope profiles. For instance, this can lead to a mass-loss runaway phenomenon and an exponential increase in the Eddington parameter, depending on the absolute mass-loss rates. Consequently, inflation could occur internally, closer to the base of the massive convective envelope of the expanded RSG model, resulting in a blended inflated and expanded morphology. This process can also lower the mass threshold required for such a blended morphology.

The location in the HR diagram where such high $\Gamma_\mathrm{rad}$ values are predicted to occur throughout the entire envelope shows a clear absence of RSGs. The combination of strong radiative instabilities in the envelope and the failure of our models to regulate their $L/M$ ratios could result in strong mass-loss episodes, which could potentially explain the observed cutoff in the luminosity distribution of RSGs.

At low metallicity, the effect of inflation across the Fe-bump is weaker. Coupled with low core-boundary mixing, this results in suppressed radius increases during both the expansion and inflation phases. The small core-to-total mass ratio favours a BSG configuration. Consequently, the model remains a BSG throughout its core-He-burning phase, avoiding the transition to a luminous RSG and potentially avoiding strong mass loss or eruptive events. The smaller core and compact envelope of BSGs suggest that the entire envelope is likely to collapse into the BH at the end of evolution. Similarly, such configurations can arise from post-collision products where H-rich material is added to the primary's envelope while the core mass remains unchanged. These configurations are among the most promising candidates for forming the heaviest stellar BHs in the Universe.

We also analysed {model configurations with a low-mass H-poor envelope on top of a He-burning core, and find they exhibit all the properties of an inflated model. Such configurations can be realised either from a single-star scenario with strong wind mass loss during the RSG phase, or from post-MS binary mass transfer via RLOF, which strips the envelope.}  Depending on the post-RLOF conditions, these binary products could either remain partially stripped ---even redward of the H-ZAMS--- throughout the entire core-He-burning phase, or experience gradual envelope stripping due to wind mass loss. In the latter case, the models would evolve bluewards through a phase previously referred to in the literature as the `puffed-up' phase, eventually ending up close to the He-ZAMS. During this transitory phase, the $\Gamma_\mathrm{rad}$ inside the model decreases due to the thinning of the H shell and the corresponding decline in the produced luminosity. This phase of evolution mirrors the process of inflation seen from the He-ZAMS, when H is slowly added on top of a He core. Importantly, the removal of an inflated zone in this context should be termed `deflation', since the star remains in thermal balance while $\Gamma_\mathrm{rad}$ decreases.

By differentiating between the phenomena of expansion, inflation, and their inverse processes, and thoroughly analysing the internal profiles of models undergoing these processes, we aim to encourage the community to exercise caution when introducing new terminology for existing physical phenomena. Although this work focuses on a few specific examples, we hope it underscores the value of thoroughly examining model profiles to gain a deeper understanding of the underlying mechanisms.

\begin{acknowledgements}
We thank the anonymous referee for constructive comments that helped improve the paper. We warmly thank the MESA developers for making their stellar evolution code publicly available. GNS and JSV are supported by STFC funding under grant numbers ST/V000233/1 and ST/Y001338/1.
\end{acknowledgements}

%
%

\bibliographystyle{aa}
\bibliography{References} 

\end{document}